\documentclass[aps,prb,times,twocolumn,amsmath,amssymb,superscriptaddress,floatfix,showpacs]{revtex4}

\usepackage{color}

\usepackage[dvips]{graphicx}
\usepackage{subfigure}
\usepackage{bbold}
\usepackage{verbatim}
\usepackage{float}
\usepackage{enumerate}
\usepackage{amsfonts}
\usepackage{multirow}

\begin{document}


\title{Angle-resolved NMR: quantitative theory of $^{75}$As T$_1$ relaxation rate in BaFe$_2$As$_2$}


\author{Andrew Smerald}
\affiliation{H.\ H.\ Wills Physics Laboratory, University of Bristol,  Tyndall Av, BS8--1TL, UK.}
\author{Nic Shannon}
\affiliation{H.\ H.\ Wills Physics Laboratory, University of Bristol,  Tyndall Av, BS8--1TL, UK.}


\date{\today}


\begin{abstract}  
While NMR measurements of nuclear energy spectra are routinely used to characterize the static 
properties of quantum magnets, the dynamical information locked in NMR $1/T_1$ relaxation 
rates remains notoriously difficult to interpret.
The difficulty arises from the fact that information about all possible low-energy spin excitations
of the electrons, and their coupling to the nuclear moments, is folded into a single number, $1/T_1$.
Here we develop a {\it quantitative} theory of the NMR $1/T_1$ relaxation rate in a collinear 
antiferromagnet, focusing on the specific example of BaFe$_2$As$_2$.  
One of the most striking features of magnetism in BaFe$_2$As$_2$ is a strong dependence 
of $1/T_1$ on the orientation of the applied magnetic field.
By careful analysis of the coupling between the nuclear and electronic moments, we show 
how this anisotropy arises from the ``filtering'' of spin fluctuations by the form-factor for 
transferred hyperfine interactions.   
This allows us to make convincing, quantitative, fits to experimental $1/T_1$ data for 
BaFe$_2$As$_2$, for different field orientations.
We go on to show how a quantitative, angle-dependent theory for the relaxation rate leads to new 
ways of measuring the dynamical parameters of magnetic systems, in particular the spin wave velocities.
\end{abstract}


\maketitle


\section{Introduction}


Nuclear magnetic resonance (NMR) has a long history as an experimental probe, with numerous uses 
throughout physics, chemistry and medicine.  
More than 60 years after its discovery~\cite{purcell46,bloch46}, it remains one of the most powerful 
techniques for investigating solid state systems. 
NMR spectra measurements of the nuclear energy level splitting are well understood, and provide a 
wealth of quantitative information concerning the static properties of magnetic materials\cite{slichter}.
However, the dynamic properties of these materials are more difficult to access.
While much of the relevant information is encoded in the NMR relaxation rate $1/T_1$, 
it can prove difficult to extract.


The problem with interpreting measurements of $1/T_1$ is that all possible fluctuations of the electron 
moments, as well as the details of the coupling to the nuclear moment, are folded into a single number.   
Despite many decades of study\cite{moriya56,moriya256,moriya63,beeman68,mila89,mila289,
thelen93,rigamonti98}, and some notable successes, theory has thus far not developed the level of 
sophistication 
required to fully utilise the information stored in these measurements.
In this paper we address this problem by developing a quantitative theory of the
$1/T_1$ relaxation rate in the magnetically ordered phase of BaFe$_2$As$_2$.


\begin{figure}[h]
\centering
\includegraphics[width=0.45\textwidth]{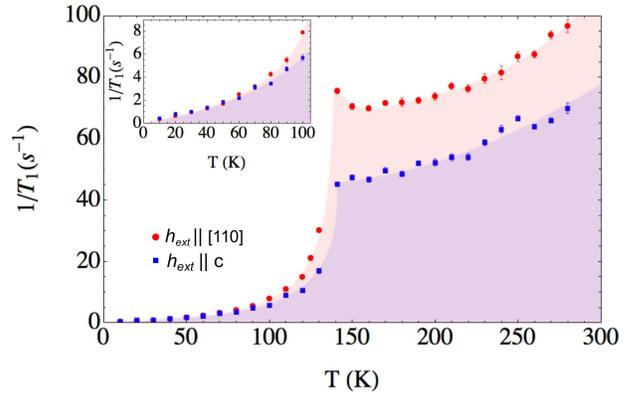}
\caption{\footnotesize{(Color online). 
Experimental measurements of the $^{75}$As $1/T_1$ relaxation rate in BaFe$_2$As$_2$, 
as reported in Ref.~[\onlinecite{kitagawa08}].   
Relaxation rates were measured with external magnetic field applied both perpendicular (squares) 
and parallel (circles) to the FeAs planes.    
Except at the lowest temperatures [inset], the relaxation rate measured for field parallel 
to $[001]$ is significantly lower than that for fields parallel to $[110]$, and has a qualitatively 
different temperature dependence.
}}
\label{fig:Ba122T1data}
\end{figure}
%


The Fe-pnictide materials in general, and BaFe$_2$As$_2$ in particular, have been a field of much activity 
in recent years\cite{kitagawa08,kitagawa09, huang08,ewings08,matan09,harriger10,ning09,smerald10,klanjsek11}.  
In BaFe$_2$As$_2$ layers of FeAs alternate with planes of Ba.  
While at room temperature the undoped materials support a tetragonal $I4/mmm$, paramagnetic phase, at 
approximately 135K there is a structural distortion to an orthorhombic $Fmmm$ phase, and subsequently 
a magnetic ordering to form a striped, collinear, antiferromagnet\cite{kitagawa08}.  
Chemical doping, such as the substitution of Co for Fe, suppresses the magnetism of the parent compounds 
and leads to an intriguing superconducting state\cite{lester09}.  


One of the many puzzling features of magnetism in BaFe$_2$As$_2$ is a significant reduction in 
the $^{75}$As $1/T_1$ NMR relaxation rate when the external magnetic field is applied perpendicular, 
rather than parallel, to the FeAs planes [Kitagawa {\it et al.}~\cite{kitagawa08}, reproduced in 
Fig.~(\ref{fig:Ba122T1data})].  
This sensitivity of $1/T_1$ to the orientation of magnetic field cannot be explained by any 
existing theory of NMR relaxation rates.


In this paper we develop a quantitative theory of the $^{75}$As NMR $1/T_1$ relaxation rate in the 
magnetically ordered phase of BaFe$_2$As$_2$, building on the earlier ideas of 
Moriya~\cite{moriya56,moriya256,moriya63}, and Mila and Rice~\cite{mila89,mila289}.   
We find that the tensor nature of transferred hyperfine interactions between electronic and 
nuclear spins leads to a ``filtering'' of the spin fluctuations seen at the $^{75}$As site, 
which in turn depends on the orientation of the magnetic field used in NMR experiments.
Consequently, the NMR $1/T_1$ relaxation rate has a {\it qualitatively different} temperature 
dependence for fields applied parallel and perpendicular to the FeAs planes.
This theory is developed in absolute units, and its predictions are found to be in {\it quantitative} 
agreement with experiment, for both orientations of the magnetic field.


In this context, it becomes meaningful to talk about ``angle-resolved'' measurements of $1/T_1$,  
and we therefore generalize our theory of $^{75}$As NMR $1/T_1$ to treat arbitrary orientation and 
magnitude of external magnetic field.
We use this more general theory to make specific, quantitative, predictions for the shape of the 
$1/T_1$ surface as a function of field orientation at fixed temperature, and the field-dependence 
of the $1/T_1$ relaxation rate as a function of field strength, at fixed field orientation.


The combination of angular resolution and absolute units also permits quantitative information 
about dynamical properties of the system to be extracted directly from $1/T_1$ measurements.
As an illustration, we show how angle-resolved $^{75}$As NMR $1/T_1$ measurements could be 
used to measure all three components of the spin wave velocity in BaFe$_2$As$_2$.  


While this paper is concerned with the specific example of $^{75}$As NMR in BaFe$_2$As$_2$, 
many of these results generalize straightforwardly to other collinear antiferromagnets, 
and corresponding results can be derived for more exotic magnetic states.
As such, angle-resolved measurement of NMR relaxation rates promises to be a powerful 
probe of both conventional and unconventional magnetism.  


The paper is structured as follows~:  
In Section~\ref{sec:background} we present some of the basic facts about antiferromagnetism in 
BaFe$_2$As$_2$, and briefly review existing theories of NMR $1/T_1$ rates in antiferromagnets.
In Section~\ref{sec:datafit}, we introduce the idea of angular resolution in $1/T_1$ measurement, 
and develop a theory for the $^{75}$As $1/T_1$ relaxation rate in the magnetically 
ordered phase of BaFe$_2$As$_2$, for fields applied in both $[110]$ and $[001]$ directions.   
In Section~\ref{sec:arbitraryfield} we show how this theory can be extended to treat 
arbitrary field orientation and strength.  
In Section~\ref{sec:characterisation}, we propose a scheme for
measuring spin wave velocities directly from NMR $1/T_1$ relaxation rates, 
based on the results of Section~\ref{sec:arbitraryfield}.
We conclude in Section~\ref{sec:conclusions} with a discussion of some of 
the wider implications of these results.


\section{Background to NMR measurements on BaFe$_2$As$_2$}
\label{sec:background}


\subsection{Low temperature magnetism in BaFe$_2$As$_2$}

\begin{figure}[h]
\centering
\includegraphics[width=0.45\textwidth]{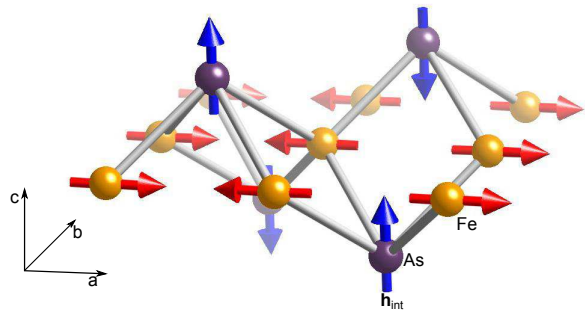}
\caption{\footnotesize{(Color online). 
Below 135K BaFe$_2$As$_2$ exhibits collinear antiferromagnet order with characteristic
wave vector \mbox{${\bf Q}=(\pi/a_0,0,\pi/c_0)$}.  
Within this ordered state Fe moments --- indicated here by red, horizontal arrows --- are orientated
along the crystalographic a-axis.   
The effective magnetic field induced at the As nucleus by these Fe moments, ${\bf h}_{\sf int}$
(blue, vertical arrows), is orientated along the crystalographic c-axis, and alternates in direction
between the As sites.
}}
\label{fig:Ba122inequivAs}
\end{figure}

Before delving into the details of the NMR relaxation rate, we briefly review the nature of the low temperature 
magnetic state in BaFe$_2$As$_2$.
Neutron scattering measurements\cite{huang08,ewings08,matan09,harriger10} reveal a 
commensurate, collinear antiferromagnetic ground state below 135K, with ordered moment 
\mbox{$m_0 \approx 0.87$$\mu_B$}~[\onlinecite{huang08}], and ordering vector 
${\bf Q}=(\pi/a_0,0,\pi/c_0)$~\cite{huang08}.   
[Here \mbox{$(a_0,b_0,c_0)=(2.80\ \AA,2.79\ \AA,6.47\ \AA)$} are the lattice 
constants of the orthorhombic lattice of Fe atoms within BaFe$_2$As$_2$]. 
The ordered magnetic moment lies along the crystallographic $a$-axis, as shown 
in Fig.~(\ref{fig:Ba122inequivAs}).   


A single branch of low-energy spin wave excitations with dispersion, 
\begin{align}
\hbar \omega_{{\bf q}} = \sqrt{\Delta^2+\sum_{\alpha=a,b,c} v_\alpha^2 (q_\alpha-Q_\alpha)^2},
\label{eq:dispersion}
 \end{align}
is found above an anisotropy gap $\Delta\approx 9.8meV$ [\onlinecite{matan09}]. 
The spin wave velocities are anisotropic with \mbox{$v_a\approx v_b \gg v_c$} 
[\onlinecite{ewings08,matan09}].    
These spin-wave excitations become diffuse at higher energies, and merge into
a broad continuum of incoherent spin excitations~\cite{matan09}.  


Constructing a theory of NMR relaxation rates is complicated by the fact that 
BaFe$_2$As$_2$ is both an antiferromagnet with a sizeable ordered moment, and a metal.
The existence of a Fermi surface in BaFe$_2$As$_2$ implies that it must support gapless 
particle-hole pairs, as well as the coherent spin-wave excitations seen in neutron scattering.
These incoherent particle-hole pairs will contribute to the $1/T_1$ relaxation rate in their own right
and, {\it a priori}, might be expected to couple strongly to spin waves.


In Section~\ref{sec:datafit} we model the magnetic excitations of BaFe$_2$As$_2$ using a low temperature 
field theory written in terms of the hydrodynamic parameters $v_a$, $v_b$, $v_c$, $\Delta$ and the 
transverse susceptibility $\chi_\perp$.  
This field theory respects all of the symmetries of the magnetic ground state, correctly reproduces the low 
temperature dispersion, Eq.~(\ref{eq:dispersion}), and provides self-consistent predictions for all  
related magnetic properties.
As such it offers a correct low-energy effective theory of spin-wave excitations in BaFe$_2$As$_2$, 
regardless of the microscopic details of the material.  


The effect of incoherent particle-hole pairs, neglected in this theory, is most evident 
in the linear-$T$ behavior of $1/T_1$ at temperatures $T < 50$$K$ 
[cf. inset to Fig.~\ref{fig:Ba122T1data}].   
We have previously argued that this contribution to $1/T_1$ remains linear in $T$ at higher 
temperatures, and can be safely fitted at low temperatures and subtracted from the 
data~\cite{smerald10}.  
We return to this argument below.   


\subsection{Introduction to NMR relaxation rates in antiferromagnets}


NMR measurements of the average internal magnetic field are well understood, and angular 
resolution is routinely used to determine the static properties of magnetic materials\cite{slichter}.
The technique relies on the Zeeman splitting of the nuclear energy levels by the effective magnetic 
field at the nuclear site, ${\bf h}_{\sf nuc}$, as described by,
\begin{align}
\mathcal{H}^\prime &= \gamma_N \hbar {\bf I}.\langle {\bf h}_{\sf nuc} \rangle  \nonumber \\
&= \gamma_N \hbar \left[ 
I^z \langle h_{\sf nuc}^z \rangle + \frac{1}{2}
 \left( I^+ \langle h_{\sf nuc}^- \rangle + I^- \langle h_{\sf nuc}^+ \rangle \right) 
 \right].
\label{eq:Hsplitting}
\end{align}
Here ${\bf I}$ is the nuclear moment, $I^{+(-)}$ is a nuclear spin raising (lowering) operator, 
and $\gamma_N$ is the gyromagnetic ratio of the nucleus in question. 


The effective magnetic field experienced by the nuclear moment  
\mbox{${\bf h}_{\sf nuc}={\bf h}_{\sf ext}+{\bf h}_{\sf int}$}, 
is the vector sum of the external magnetic field applied to the sample, ${\bf h}_{\sf ext}$, 
and an effective internal magnetic field, ${\bf h}_{\sf int}$.
This internal field encapsulates the effect of interactions 
between the nuclear moment and the surrounding electrons.  
However, while the external field ${\bf h}_{\sf ext}$ can be considered to be static, ${\bf h}_{\sf int}$ 
fluctuates on a time scale set by the electrons, and so contributes to relaxation of 
the nuclear spins.


In general, three different types of interaction can contribute to the effective 
internal field ${\bf h}_{\sf int}$.  
Firstly, for the nuclei of magnetic atoms, there is an on-site hyperfine interaction.
Secondly, for the nuclei of non-magnetic atoms, there is a transferred hyperfine 
coupling between the nuclear moment and the spin of neighbouring electrons.
Finally, we can also consider the dipolar interaction between nuclear 
and electronic spins.
This is weak for small values of electronic and/or nuclear spin, but long-ranged.
In the case of NMR in BaFe$_2$As$_2$, we focus in particular 
on the transferred hyperfine coupling between an $^{75}$As nucleus, 
and the electrons of the four Fe atoms which surround it. 


In NMR $1/T_1$ measurements, the population of the Zeeman split nuclear energy 
levels is driven out of equilibrium by a radio-frequency pulse.  
These nuclear spins then return to equilibrium over a characteristic timescale $T_1$, 
which is set by their interaction with electrons --- specifically by the transverse fields 
$h^+$ and $h^-$ in Eq.~(\ref{eq:Hsplitting}).  
We characterize this process by the standard $1/T_1=2W$ relaxation rate, as defined in 
[\onlinecite{andrew61,narath67,watanabe94}]. 


The quest for a theory of NMR $1/T_1$ relaxation rates in antiferromagnetic materials 
now spans almost six decades.   
Moriya, writing in 1956, was the first to realise how the Raman scattering of antiferromagnetic spin waves 
from nuclear moments can lead to a finite nuclear spin relaxation rate\cite{moriya56,moriya256,moriya63}.    
Beeman and Pincus later extended Moriya's semi-classical theory to include quantum 
fluctuation effects\cite{beeman68}. 
The next major breakthrough was due to Mila and Rice, who realised that the indirect coupling 
between electronic spins and the nuclei of non-magnetic atoms can act as a ``filter'' on spin 
fluctuations\cite{mila89,mila289}.
This made it possible to understand for the first time why the $1/T_1$ relaxation rates of different 
nuclei in the same compound can have qualitatively different leading temperature dependences.


The theory of NMR relaxation rate presented in this paper extends Mila and Rice's idea of 
``filtering'' of spin fluctuations by showing how the action of the ``filter'' is strongly dependent 
on the orientation of the externally applied magnetic field. 
We develop this theory in absolute units, which allows us to make quantitative predictions for 
comparison with experiment.
Using this theory, we are able explain the temperature dependence of $^{75}$As relaxation rates 
in BaFe$_2$As$_2$, for fields applied in both the $[110]$ and $[001]$ directions~\cite{kitagawa08}.   
This task is simplified by the very large anisotropy gap $\Delta \approx 100K$ in the spin wave 
spectrum of BaFe$_2$As$_2$, which forbids additional relaxation processes arising from the 
absorption or emission of spin waves\cite{moriya56,beeman68}.


We note that a theory of NMR  $1/T_1$ relaxation rates in magnetic Fe pnictides has also 
been advanced by Ong {\it et al.}~\cite{ong09}. 
However the theory we present goes much further and allows quantitative comparison with experiment~\cite{kitagawa08}. 


We also remark that BaFe$_2$As$_2$ is not the first material in which  the NMR $1/T_1$ 
relaxation rate has been found to depend on the direction in which the magnetic field was applied 
--- although it is, to the best of our knowledge, the first antiferromagnet.
Angle dependence of $1/T_1$ relaxation rates has previously been observed in $^{63}$Cu 
NMR of the cuprates YBa$_2$Cu$_3$O$_7$ and YBa$_2$Cu$_4$O$_8$, 
within their low-temperature superconducting state\cite{takigawa91,bankay92}.   
Thelen, Pines and Lu\cite{thelen93} have suggested that the anisotropy in $1/T_1$
follows from an anisotropy in the on-site coupling between the electron and nuclear 
moments of $^{63}$Cu.  
In this paper we focus on a non-magnetic ion, $^{75}$As, for which there is no on-site coupling, 
and explain the angle dependence in terms of transferred hyperfine interactions.


\section{Quantitative theory of $^{75}$As $1/T_1$ for field parallel to $[110]$ and $[001]$}
\label{sec:datafit}


We now develop a theory of $1/T_1$ relaxation rates in BaFe$_2$As$_2$, with the specific 
goal of obtaining quantitative fits to high-quality $^{75}$As NMR data for magnetic fields 
parallel to both the $[110]$ and $[001]$ directions~\cite{kitagawa08}.  
We focus on results for the relaxation rate at low temperatures and in the magnetically ordered phase, 
and follow a relatively direct path to the results needed to compare with experiment.   
The development of a more general theory for arbitrary field orientation is postponed to 
Section~\ref{sec:arbitraryfield}.


If the nuclear field, ${\bf h}_{\sf nuc}$, is orientated in the $z$-direction, then time dependent 
perturbation theory leads to\cite{moriya63},
\begin{align}
\frac{1}{T_1({\bf h}_{\sf ext})} = \frac{\gamma_N^2}{2} \int dt \ e^{i\omega_0 t} 
& \left[ \left\langle \left\{  h^x_{\sf int}(t),h^x_{\sf int}(0) \right\} \right\rangle \right. \nonumber \\
& \left. + \left\langle \left\{  h^y_{\sf int}(t),h^y_{\sf int}(0) \right\} \right\rangle  \right],
\label{eq:linresponse}
\end{align}
where $\hbar \omega_0$ is the splitting of the nuclear energy levels and ${\bf h}_{\sf int}$ is the 
effective field at the nuclear site due to interaction with the surrounding electron moments \footnote{In reference [\onlinecite{smerald10}] we used the symbol, $\mathcal{B}$, to denote the nuclear-electron coupling tensor. In this paper we instead use $\mathcal{A}$, since this appears to be a more widely accepted nomenclature.}.
Only the $x$ and $y$ components enter the expression, since they are the ones that couple to 
the nuclear spin raising and lowering operators, $I^\pm$, in Eq.~(\ref{eq:Hsplitting}).


The internal field, ${\bf h}_{\sf int}$, can be re-expressed in terms of the electronic degrees of freedom as,
\begin{eqnarray}
{\bf h}_{\sf int}(t) = \sum_i \underline{\underline{\mathcal{A}}}_i . {\bf m}_i(t),
\label{eq:intfield}
\end{eqnarray}
where $i$ sums over the electron moments (${\bf m}_i$) that couple to the nuclear spin, and 
$\underline{\underline{\mathcal{A}}}_i$ $[T/\mu_B]$ is a rank two tensor describing the coupling 
between the nuclear and electron moments.  
We refer to $\underline{\underline{\mathcal{A}}}_i$as the nuclear-electron coupling tensor, 
and note that its components,
\begin{align}
\underline{\underline{\mathcal{A}}}_i=
\left(
\begin{array}{ccc}
\mathcal{A}_i^{11} &\mathcal{A}_i^{12} & \mathcal{A}_i^{13} \\
\mathcal{A}_i^{21} &\mathcal{A}_i^{22} & \mathcal{A}_i^{23} \\
\mathcal{A}_i^{31} & \mathcal{A}_i^{32} & \mathcal{A}_i^{33} 
\end{array}
\right),
\label{eq:Aitensor}
\end{align}
 can be measured by Knight shift experiments\cite{kitagawa08}.  
 The full interaction Hamiltonian for a nuclear moment, ${\bf I}$, is given by,
\begin{align}
\mathcal{H} = \gamma_N \hbar \left( {\bf I}.{\bf h}_{\sf ext} + \sum_i {\bf I}.\underline{\underline{\mathcal{A}}}_i . {\bf m}_i(t)  \right).
\end{align} 


\begin{figure}[h]
\centering
\includegraphics[width=0.45\textwidth]{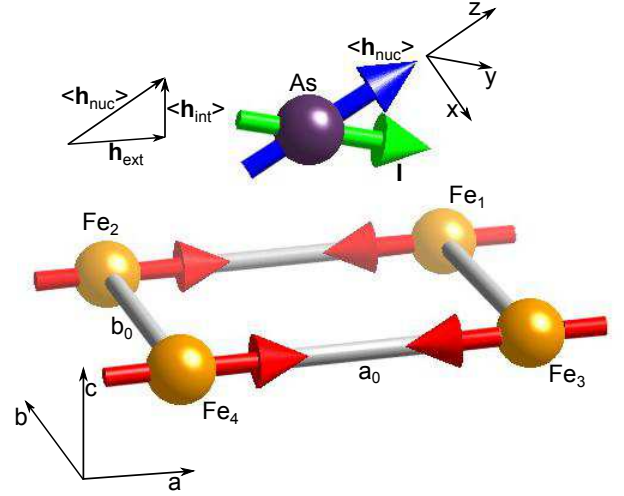}
\caption{\footnotesize{(Color online).  
The local environment of the As atom in BaFe$_2$As$_2$.  
The As atom (purple) experiences an average local field $\langle {\bf h}_{\sf nuc} \rangle$ (blue arrow) 
that arises from a combination of an externally applied field ${\bf h}_{\sf ext}$, shown here in the $a$-direction, 
and an internal field $\langle {\bf h}_{\sf int} \rangle$.  
The internal field points on average in the $c$-direction, and is due to the interaction with the electron 
magnetic moments (average position shown by red arrows).  
After being excited by a radiation pulse, the As nuclear moment ${\bf I}$ (green arrow) relaxes back towards 
alignment with the magnetic field.
}}
\label{fig:Ba122plaquette}
\end{figure}


Starting from Eq.~(\ref{eq:linresponse}), we now show how to derive a working expression for the 
relaxation rate by making use of the fluctuation dissipation theorem.  
Eq.~(\ref{eq:intfield}) and  Eq.~(\ref{eq:linresponse}) can be combined to give,
\begin{align}
\frac{1}{T_1({\bf h}_{\sf ext})} = &\frac{\gamma_N^2}{2} \int dt \ e^{i\omega_0 t} \frac{1}{N} \sum_{\bf q} \sum_{i,j}  e^{i{\bf q}.({\bf r}_i-{\bf r}_j)}  \nonumber \\
&  \left[   \left\langle   \{  ( \underline{\underline{\mathcal{A}}}_i . {\bf m}_{{\bf q}}(t)  )^x, ( \underline{\underline{\mathcal{A}}}_i . {\bf m}_{-{\bf q}}  )^x \} \right\rangle \right. \nonumber \\
&\left. + \left\langle   \{  ( \underline{\underline{\mathcal{A}}}_i . {\bf m}_{{\bf q}}(t)  )^y, ( \underline{\underline{\mathcal{A}}}_i . {\bf m}_{-{\bf q}}  )^y \} \right\rangle   \right],
 \label{eq:hyperfinerelaxation}
\end{align}
where the electron magnetic moments have been Fourier transformed using,
\begin{align}
 {\bf m}_i(t)= \frac{1}{\sqrt{N}} \sum_{\bf q} e^{i{\bf q}.{\bf r}_i}   {\bf m}_{\bf q}(t).
\end{align}
Fluctuations of the electronic moments can be characterised by introducing a dynamical structure factor,
\begin{eqnarray}
\mathcal{S}^{\xi\psi}({\bf q},\omega_0) = \int dt \ e^{i\omega_0 t}  \left\langle   \{  m^\xi_{{\bf q}}(t), m^\psi_{-{\bf q}} \}  \right\rangle.
\end{eqnarray}
It follows that the relaxation rate is given by,
\begin{align}
\frac{1}{T_1({\bf h}_{\sf ext})} = \frac{\gamma_N^2}{2N}   \sum_{{\bf q},\xi,\psi} 
\left[ \mathcal{A}_{\bf q}^{x\xi}\mathcal{A}_{-{\bf q}}^{x\psi} + \mathcal{A}_{\bf q}^{y\xi}\mathcal{A}_{-{\bf q}}^{y\psi} \right]
\mathcal{S}^{\xi\psi}({\bf q},\omega_0),
\label{eq:strucfac}
\end{align}
where $\xi,\psi=\{ x,y,z\}$ and we have defined,
\begin{align}
\mathcal{A}_{\bf q}^{\xi\psi}=\sum_i e^{i{\bf q}.{\bf r}_i} \mathcal{A}_i^{\xi\psi}.
\label{eq:AFT}
\end{align}


The fluctuation-dissipation theorem relates the structure factor of Eq.~(\ref{eq:strucfac}) to the dynamic 
susceptibility, defined by,
\begin{align}
\chi^{\xi\psi}({\bf q},\omega_0) = i \int dt \ e^{i\omega_0 t}  \left\langle   [  \delta m^\xi_{{\bf q}}(t), \delta m^\psi_{-{\bf q}} ]  \right\rangle.
\label{eq:suscepdefinition}
\end{align}
This allows the relaxation rate to be rewritten as,
\begin{align}
&\frac{1}{T_1({\bf h}_{\sf ext})} =  \lim_{\omega_0\rightarrow 0}\frac{\gamma_N^2}{2N}  k_BT \nonumber \\
&\times \sum_{{\bf q},\xi,\psi} 
\left[ \mathcal{A}_{\bf q}^{x\xi}\mathcal{A}_{-{\bf q}}^{x\psi} + \mathcal{A}_{\bf q}^{y\xi}\mathcal{A}_{-{\bf q}}^{y\psi} \right]
\frac{\Im m \left\{ \chi^{\xi\psi}({\bf q},\omega_0) \right\} }{\hbar \omega_0},
\label{eq:T1suscep}
\end{align}
where, in taking the limit $\omega_0\rightarrow 0$, we have assumed that the energy of the nuclear 
transitions is negligible compared to the typical spin wave energies. 


Since both the susceptibility, Eq.~(\ref{eq:suscepdefinition}), and the interaction between the nuclear 
and electron moments, Eq.~(\ref{eq:Aitensor}), are tensors, it is important to keep track of the coordinate 
basis in which they are represented.
In the above expressions, both are represented in the $(x,y,z)$ coordinate system, which is defined 
by aligning the $z$-axis with ${\bf h}_{\sf nuc}$, the nuclear magnetic field.  
However, they are most simply measured and calculated in the $(a,b,c)$ coordinate system, which is aligned 
with the crystal axes.  
Fig.~(\ref{fig:Ba122plaquette}) shows the orientation of these two coordinate systems when ${\bf h}_{\sf ext}$ 
is parallel to the $a$-axis.  
In Section~\ref{sec:arbitraryfield} we consider an arbitrary direction of  ${\bf h}_{\sf ext}$, and introduce 
rotation matrices in order to map between the $(x,y,z)$ and $(a,b,c)$ bases. 
Here we specialise to the two cases measured by Kitagawa {\it et al}: ${\bf h}_{\sf ext}$ orientated in the 
$[110]$ direction; and ${\bf h}_{\sf ext}$ parallel to the $c$-axis. 


In what follows, we make the assumption \mbox{$|{\bf h}_{\sf ext}|\gg|{\bf h}_{\sf int}|$}. 
This means that \mbox{${\bf h}_{\sf nuc}\approx {\bf h}_{\sf ext}$} and the $z$-axis is aligned with the 
external magnetic field.  
The discussion of an arbitrary magnitude of external field is postponed to Section~\ref{sec:arbitraryfield}.
 If ${\bf h}_{\sf ext}$ is applied in the $c$-direction then the $(x,y,z)$ and $(a,b,c)$  coordinate systems 
 are equivalent. In consequence, the tensors appearing in Eq.~(\ref{eq:T1suscep}) can be transformed 
 into the $(a,b,c)$ coordinate system by  the substitution \mbox{$(x\rightarrow a,y\rightarrow b,z\rightarrow c)$}.
 If instead ${\bf h}_{\sf ext}$ is applied along the $[110]$-direction, a set of rotation matrices is required 
 to relate the tensors expressed in the two  coordinate systems.  
 Again we postpone the details of this to Section~\ref{sec:arbitraryfield} but, at a schematic level, we 
 make the substitution  \mbox{$(x\rightarrow (b-a)/\sqrt{2},y\rightarrow c,z\rightarrow (a+b)/\sqrt{2})$}.


We have already argued that, due to the sizeable gap in the spin-wave spectrum, the relaxation is dominated 
by the scattering of  spin wave excitations. 
This corresponds to picking out the longitudinal component of the susceptibility tensor in the crystallographic 
$(a,b,c)$ coordinate system.  
For collinear magnetic order in the $a$-direction this is the component 
$\chi^{aa}({\bf q},\omega_0)=\chi_\parallel({\bf q},\omega_0)$. 
Thus  the relaxation rate can be expressed as,
 \begin{align}
\frac{1}{T_1({\bf h}_{\sf ext})} 
   &=  \lim_{\omega_0\rightarrow 0}\frac{\gamma_N^2}{2N}  k_BT \sum_{{\bf q}\in \mathrm{PMBZ}} \nonumber \\
   & \mathcal{F}({\bf q},{\bf h}_{\sf ext})
\frac{ \Im m \left\{ \chi_\parallel({\bf q},\omega_0) \right\} }{\hbar \omega_0},
\label{eq:T1general1}
\end{align}
where $\mathcal{F}({\bf q},{\bf h}_{\sf ext})$ is a form factor for the nuclear-electron interaction, 
which acts as the ``filter'' of spin fluctuations.
The sum is over all ${\bf q}$ vectors in the Fe-paramagnetic Brillouin Zone (PMBZ).
For external field in the $[110]$-direction,
 \begin{align}
\mathcal{F}({\bf q}, h^{110}_{\sf ext})= \frac{1}{2}\left( \mathcal{A}_{\bf q}^{aa} -\mathcal{A}_{\bf q}^{ba} \right) \left( \mathcal{A}_{-{\bf q}}^{aa} -\mathcal{A}_{-{\bf q}}^{ba} \right) +\mathcal{A}_{\bf q}^{ca}\mathcal{A}_{-{\bf q}}^{ca},
\label{eq:abformfactorgeneral}
\end{align}
while for external field in the $c$-direction,
 \begin{align}
\mathcal{F}({\bf q}, h^{c}_{\sf ext})= \mathcal{A}_{\bf q}^{aa}\mathcal{A}_{-{\bf q}}^{aa} + \mathcal{A}_{\bf q}^{ba}\mathcal{A}_{-{\bf q}}^{ba}.
\label{eq:cformfactorgeneral}
\end{align}


Eq.~(\ref{eq:T1general1}) gives the relaxation rate in terms an integral over the product of the imaginary 
part of the electronic, dynamic susceptibility, and a form factor that encapsulates the interaction between 
the electronic and nuclear moments. 
These two quantities can be developed independently, and then recombined to find the $1/T_1$ relaxation rate.


\subsection{Form factor}


In this subsection we determine the form factors necessary to explain the experimental data
shown in Fig.~(\ref{fig:Ba122T1data}) [\onlinecite{kitagawa08}].  
This requires an appreciation of the symmetry of the nuclear environment, and below 
we outline an analysis of the nuclear-electron coupling tensor similar to that carried 
out in [\onlinecite{kitagawa08}].


For the case of the As atom in BaFe$_2$As$_2$, the dominant interaction\cite{kitagawa08} with the electron system  is via a transferred hyperfine coupling with the four nearest neighbour Fe electron moments, shown in Fig.~(\ref{fig:Ba122plaquette}).


The environment of the As nuclear moment is thus invariant under the symmetry operations of the point group $C_{2v}$. Referring to the labelling of the Fe moments shown in Fig.~(\ref{fig:Ba122plaquette}), the nuclear-electron coupling tensor for the first Fe site can be written as,
\begin{align}
\underline{\underline{\mathcal{A}}}_1=
\left(
\begin{array}{ccc}
\mathcal{A}^{aa} &\mathcal{A}^{ab} & \mathcal{A}^{ac} \\
\mathcal{A}^{ba} &\mathcal{A}^{bb} & \mathcal{A}^{bc} \\
\mathcal{A}^{ca} & \mathcal{A}^{cb} & \mathcal{A}^{cc} 
\end{array}
\right).
\end{align}
Reflection symmetry in the $bc$-plane allows the tensor for the second Fe site to be determined as,
\begin{align}
\underline{\underline{\mathcal{A}}}_2=
\left(
\begin{array}{ccc}
\mathcal{A}^{aa} &-\mathcal{A}^{ab} & -\mathcal{A}^{ac} \\
-\mathcal{A}^{ba} &\mathcal{A}^{bb} & \mathcal{A}^{bc} \\
-\mathcal{A}^{ca} & \mathcal{A}^{cb} & \mathcal{A}^{cc} 
\end{array}
\right).
\end{align}
By reflection in the $ac$-plane, 
\begin{align}
\underline{\underline{\mathcal{A}}}_3=
\left(
\begin{array}{ccc}
\mathcal{A}^{aa} &-\mathcal{A}^{ab} & \mathcal{A}^{ac} \\
-\mathcal{A}^{ba} &\mathcal{A}^{bb} & -\mathcal{A}^{bc} \\
\mathcal{A}^{ca} & -\mathcal{A}^{cb} & \mathcal{A}^{cc} 
\end{array}
\right),  
\end{align}
and a by $\pi$ rotation around the $c$-axis,
\begin{align}
\underline{\underline{\mathcal{A}}}_4=
\left(
\begin{array}{ccc}
\mathcal{A}^{aa} &\mathcal{A}^{ab} & -\mathcal{A}^{ac} \\
\mathcal{A}^{ba} &\mathcal{A}^{bb} & -\mathcal{A}^{bc} \\
-\mathcal{A}^{ca} & -\mathcal{A}^{cb} & \mathcal{A}^{cc} 
\end{array}
\right).  
\end{align}
These four tensors can be combined using Eq.~(\ref{eq:AFT}) to find,
\begin{align}
\underline{\underline{\mathcal{A}}}_{\bf q}= 
4\left(
\begin{array}{ccc}
\mathcal{A}^{aa}c_a  c_b & -\mathcal{A}^{ab} s_a s_b & i \mathcal{A}^{ac} s_a c_b \\
-\mathcal{A}^{ba}s_a s_b &\mathcal{A}^{bb} c_a  c_b & i\mathcal{A}^{bc}c_a s_b \\
i\mathcal{A}^{ca}  s_a c_b & i\mathcal{A}^{cb}c_a s_b & \mathcal{A}^{cc} c_a  c_b
\end{array}
\right), 
\label{eq:Aqmatrix} 
\end{align}  
where,
\begin{align}
&c_a =\cos \frac{q_aa_0}{2}, \qquad c_b =\cos \frac{q_bb_0}{2}  \nonumber \\
&s_a=\sin \frac{q_aa_0}{2}, \qquad s_b =\sin \frac{q_bb_0}{2}.
\label{eq:casa}
\end{align}  


The internal field follows from substituting the above expressions for the nuclear-electron coupling tensors into Eq.~(\ref{eq:intfield}). We find, in agreement with [\onlinecite{kitagawa08}], an average internal field along the $c$-direction given by,
\begin{align}
\langle {\bf h}_{\sf int} \rangle=
\pm 4\mathcal{A}^{ca} \langle m^a \rangle \left(
\begin{array}{c}
0 \\
0 \\
1
\end{array}
\right),
\label{eq:hintc}
\end{align}
where $\langle m^a \rangle$ is the component of the average, electronic magnetic moment 
aligned with the $a$-axis. 
The sign of the field depends on which As nucleus is under investigation.


\begin{figure}[ht]
\centering
\includegraphics[width=0.45\textwidth]{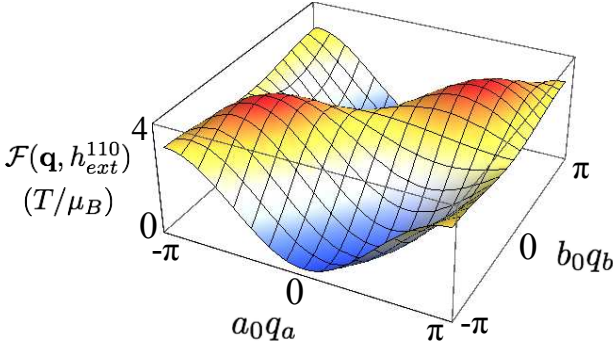}
\caption{\footnotesize{(Color online).  
${\bf q}$-dependence of the form factor $\mathcal{F}({\bf q}, h^{110}_{\sf ext})$, given in 
Eq.~(\ref{eq:abformfactorallq}), for an external field  applied in the $[110]$-direction. 
The form factor acts as a ``filter'' of the electronic fluctuations. 
It is finite at the ordering vector ${\bf q}=(\pi/a_0,0,\pi/c_0)$, and therefore allows the dominant 
fluctuations of the longitudinal susceptibility to ``pass the filter''.  
We use the parameters $\mathcal{A}^{aa}=0.66 \ T/\mu_B$ and 
\mbox{$\mathcal{A}^{ca}=0.43 \ T/\mu_B$} from [\onlinecite{kitagawa08}] and 
make the approximation \mbox{$\mathcal{A}^{aa}\approx\mathcal{A}^{ba}$}. 
}}
\label{fig:formfactorab}
\end{figure}


Eq.~(\ref{eq:abformfactorgeneral}) and Eq.~(\ref{eq:Aqmatrix}) can be used to calculate the form 
factor for field applied in the $[110]$-direction as,
\begin{align}
\mathcal{F}({\bf q}, h^{110}_{\sf ext})= 8\left( \mathcal{A}^{aa}c_ac_b 
   +\mathcal{A}^{ba}s_as_b \right)^2  +16(\mathcal{A}^{ca}s_ac_b)^2,
\label{eq:abformfactorallq}
\end{align}
where, following Eq.~(\ref{eq:casa}), $c_a$, $c_b$, $s_a$ and $s_b$ are ${\bf q}$-dependent. 
For this field orientation, the form factor is finite at the ordering vector 
\mbox{${\bf q}={\bf Q}=(\pi/a_0,0,\pi/c_0)$}, as shown in  Fig.~(\ref{fig:formfactorab}). 
Exactly at this point \mbox{$\mathcal{F}({\bf Q}, h^{110}_{\sf ext})=16(\mathcal{A}^{ca})^2$}, and this 
is the approximation used in reference [\onlinecite{smerald10}]. 


\begin{figure}[h]
\centering
\includegraphics[width=0.45\textwidth]{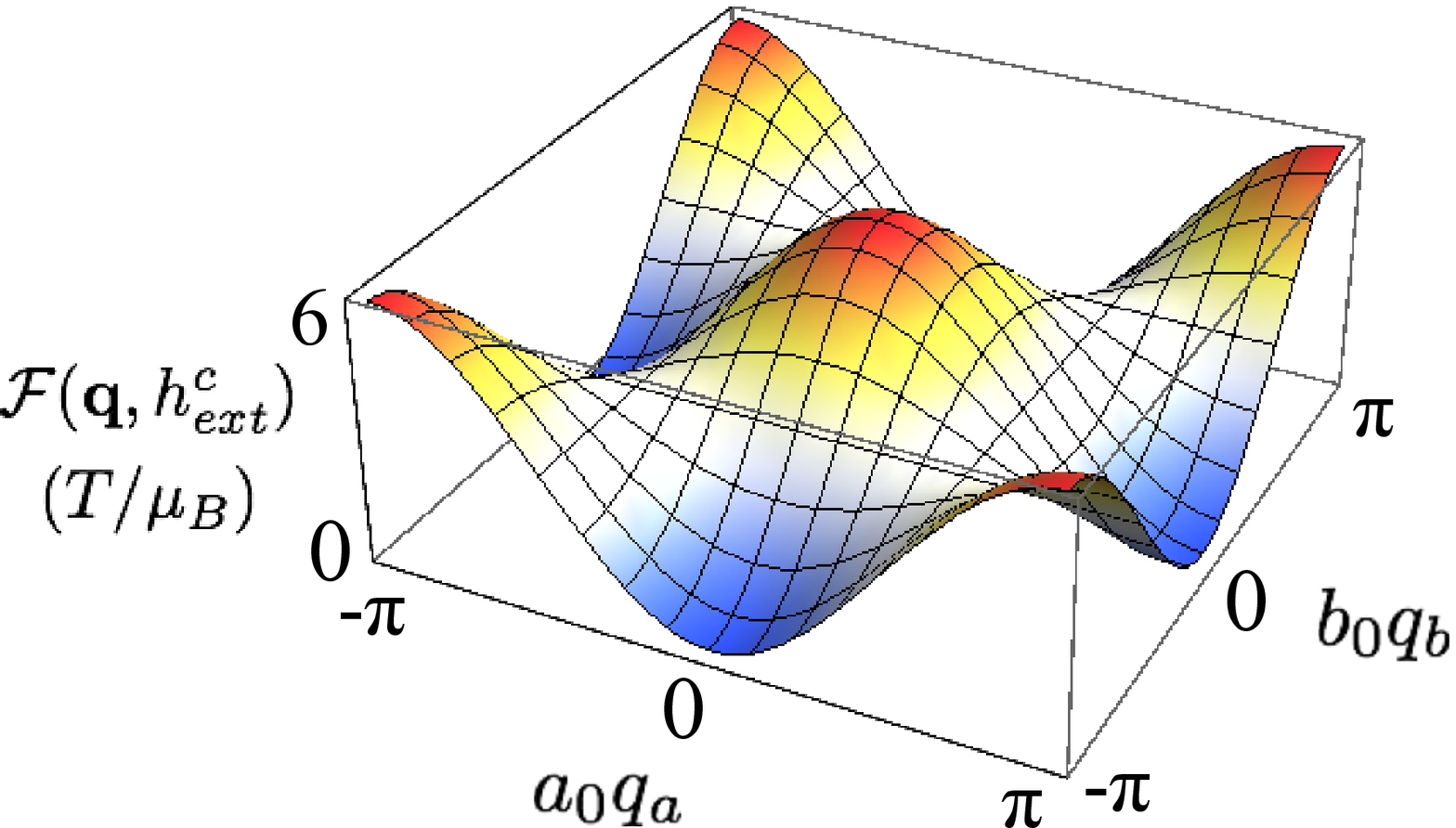}
\caption{\footnotesize{(Color online).  
${\bf q}$-dependence of the form factor $\mathcal{F}({\bf q}, h^{c}_{\sf ext})$, 
given in Eq.~(\ref{eq:cformfactorallq}), for an external field  applied in the $c$-direction. 
The form factor is zero at \mbox{${\bf q}=(\pi/a_0,0,q_c)$}, and therefore it ``filters out'' the dominant 
electronic fluctuations at \mbox{${\bf q}=(\pi/a_0,0,\pi/c_0)$}.  
It has a maximum at \mbox{${\bf q}=(0,0,0)$}, which matches the secondary peak in the imaginary 
part of the longitudinal susceptibility.  
We use the parameter \mbox{$\mathcal{A}^{aa}=0.66 \ T/\mu_B$}  from [\onlinecite{kitagawa08}] 
and make the approximation $\mathcal{A}^{aa}\approx\mathcal{A}^{ba}$.
}}
\label{fig:formfactorc}
\end{figure}
For field applied in the $c$-direction Eq.~(\ref{eq:cformfactorgeneral}) and Eq.~(\ref{eq:Aqmatrix}) 
can be combined to find,
\begin{align}
\mathcal{F}({\bf q}, h^{c}_{\sf ext})= 16 (\mathcal{A}^{aa}c_ac_b)^2 +16(\mathcal{A}^{ba}s_as_b)^2. 
\label{eq:cformfactorallq}
\end{align}


This has a peak at \mbox{${\bf q}=(0,0,0)$} and is zero at \mbox{${\bf q}=(\pi/a_0,0,\pi/c_0)$}.  
Thus fluctuations of the electron system at the ordering vector \mbox{${\bf Q}$} will be ``filtered out'' by the form factor (see Fig.~(\ref{fig:formfactorc})).


In principle one could also include a  dipolar coupling to the surrounding electron moments.  
The relevant nuclear-electron coupling tensor is given by,
\begin{align}
\underline{\underline{\mathcal{A}}}^{dip}_i=
-\frac{\mu_0}{4\pi r_i^3}
\left(
\begin{array}{ccc}
1-3a_i^2/r_i^2 & -3a_ib_i/r_i^2 & -3a_ic_i/r_i^2 \\
-3b_ia_i/r_i^2 &1-3b_i^2/r_i^2 & -3b_ic_i/r_i^2 \\
-3c_ia_i/r_i^2 & -3c_ib_i/r_i^2 & 1-3c_i^2/r_i^2
\end{array}
\right),
\label{Eq:dipolartensor}
\end{align}
where ${\bf r}_i=(a_i,b_i,c_i)$ is a vector connecting the $i$th electron moment to the nuclear site.  
This is longer range than the hyperfine coupling, but the  symmetry of the nuclear environment 
remains $C_{2v}$.  
In consequence the position of the peaks and troughs in the form factor are unchanged, 
and thus the qualitative structure of the relaxation rate will be the same.  
Since this form of coupling has been shown to be negligible in BaFe$_2$As$_2$ 
[\onlinecite{kitagawa08}], we concentrate exclusively on the hyperfine interaction.


\subsection{Dynamical, longitudinal susceptibility}
\label{subsec:suscep}


We now turn to  the dynamical susceptibility of the electron moments, concentrating on the longitudinal 
fluctuations relevant to BaFe$_2$As$_2$.  
The low energy field theory that reproduces the dispersion relation and has all the correct symmetries 
of the ordered state is\cite{chakravarty89,allen97,benfatto06,ong09,smerald10},
 \begin{align}
\label{eq:nlsm} 
\mathcal{S}[{\bf n}] &=\frac{1}{2\hbar V_{cell}} \int d^3r d\tau 
[  
\hbar^2\chi_\perp (\partial_\tau {\bf n})^2 + \hspace{-2mm} \sum_{\alpha=a,b,c} \hspace{-2mm} \rho_\alpha (\partial_\alpha {\bf n})^2 
\nonumber\\
& \qquad\qquad\qquad \qquad \qquad \qquad 
  -\chi_\perp \Delta^2 n_a^2  
 ],
\end{align} 
where $\chi_\perp$ is the static perpendicular susceptibility and $\rho_\alpha$ is the spin stiffness along 
the $\alpha$th crystallographic direction. 
The relation between spin stiffness and spin wave velocity is \mbox{$v_\alpha = \sqrt{\rho_\alpha/\chi_\perp}$}.  
The action is based on the non-linear sigma model, whose non-linearity arises from the requirement that 
${\bf n}^2=1$ in the partition function,
\begin{align}
\mathcal{Z} = \int \mathcal{D} {\bf n} \ \delta({\bf n}^2-1) \ e^{-\mathcal{S}[{\bf n}] }.
\end{align}
The constraint, ${\bf n}^2=1$, remains valid in the anisotropic model, Eq.~(\ref{eq:nlsm}), and the anisotropic term, 
\mbox{$\chi_\perp \Delta^2 n_a^2$}, arises from the fact that the $a$-axis is energetically favoured as the ordering 
direction in BaFe$_2$As$_2$.
While the correct microscopic model for electronic magnetism in the pnictide materials remains 
controversial\cite{yildrim08,si08,yi09,eremin10,kaneshita10,knolle10,schmidt10}, we stress that 
this field theory provides a correct description of their low-energy spin-wave excitations, regardless of the
details of the high-energy physics. 


The longitudinal, dynamic susceptibility follows from the action given in Eq.~(\ref{eq:nlsm}).  
In Appendix A we derive an expression for the susceptibility, using a Gaussian approximation 
to describe fluctuations of the order-parameter field,~${\bf n}$, around the ordered state.
This has two main contributions, one from ${\bf q}\approx {\bf Q}$ and the other from ${\bf q}\approx 0$, 
and can be expressed as,
\begin{align}
\Im m \left\{ \chi_{\parallel}({\bf q},\omega_0) \right\} \approx &
\Im m \left\{ \chi_{\parallel,st}({\bf q}\approx {\bf Q},\omega_0) \right\} \nonumber \\
&+\Im m \left\{ \chi_{\parallel,un}({\bf q}\approx 0,\omega_0) \right\}.
\label{eq:suscepsum}
  \end{align}
Taking the limit $\omega_0 \rightarrow 0 $ in Eq.~(\ref{eqapp:stagsuscep}) and 
Eq.~(\ref{eqapp:unsuscep}) gives,
\begin{align}
& \Im m \left\{ \chi_{\parallel,st}({\bf q}\approx {\bf Q} ,\omega_0) \right\} \approx 
 \left(\frac{g_l\mu_BS}{2} \right)^2 \pi \hbar V_{cell} \frac{\hbar \omega_0}{k_BT} \frac{1}{\chi_\perp^2} 
 \nonumber \\
&  \left\lgroup \int_{{\bf k}\approx 0} \frac{d^3k}{(2\pi)^3}   \frac{n_B(\omega_{1,{\bf k}})(n_B(\omega_{1,{\bf k}})+1)}{(\hbar\omega_{1,{\bf k}})^2} \right. \nonumber \\
& \qquad \qquad \qquad \qquad \qquad \qquad \delta(\hbar\omega_{1,{\bf k}}-\hbar\omega_{2,{\bf k}+{\bf q}}) \nonumber \\
&    + \int_{{\bf k}\approx {\bf Q}} \frac{d^3k}{(2\pi)^3}\frac{n_B(\omega_{2,{\bf k}})(n_B(\omega_{2,{\bf k}})+1)}{(\hbar\omega_{2,{\bf k}})^2 }    \nonumber \\
& \left.  \qquad \qquad \qquad \qquad \qquad \qquad \delta(\hbar\omega_{2,{\bf k}}-\hbar\omega_{1,{\bf k}+{\bf q}}) \right\rgroup,  \nonumber \\
\label{eq:staggeredfieldsuscep}
\end{align}
and,
\begin{align}
& \Im m \left\{ \chi_{\parallel,un}({\bf q}\approx 0,\omega_0) \right\} \approx
(g_l\mu_B)^2  \pi \hbar V_{cell} \frac{\hbar \omega_0}{k_BT}  
 \nonumber \\
&  \left\lgroup 
 \int_{{\bf k}\approx 0} \frac{d^3k}{(2\pi)^3}  n_B(\omega_{1,{\bf k}})(n_B(\omega_{1,{\bf k}})+1) \delta(\hbar\omega_{1,{\bf k}}-\hbar\omega_{1,{\bf k}+{\bf q}})  \right. \nonumber \\
& \left.  + \int_{{\bf k}\approx {\bf Q}}  \frac{d^3k}{(2\pi)^3}  n_B(\omega_{2,{\bf k}})(n_B(\omega_{2,{\bf k}})+1) \delta(\hbar\omega_{2,{\bf k}+{\bf q}}-\hbar\omega_{2,{\bf k}}) 
\right\rgroup,
\label{eq:constfieldsuscep} 
  \end{align}  
where $S$ is the average electron spin per Fe-site, $g_l$ is the land\'{e} g-factor, $n_B$ 
is the standard Bose factor and,
\begin{align}
&\hbar \omega_{1,{\bf q}} = \sqrt{\Delta^2+\sum_\alpha v_\alpha^2 q_\alpha^2} \nonumber \\
&\hbar \omega_{2,{\bf q}} = \sqrt{\Delta^2+\sum_\alpha v_\alpha^2 (q_\alpha-Q_\alpha)^2}.
\label{eq:dispersion12}
 \end{align}

\begin{figure}[ht]
\centering
\subfigure[]{\includegraphics[width=0.45\textwidth]{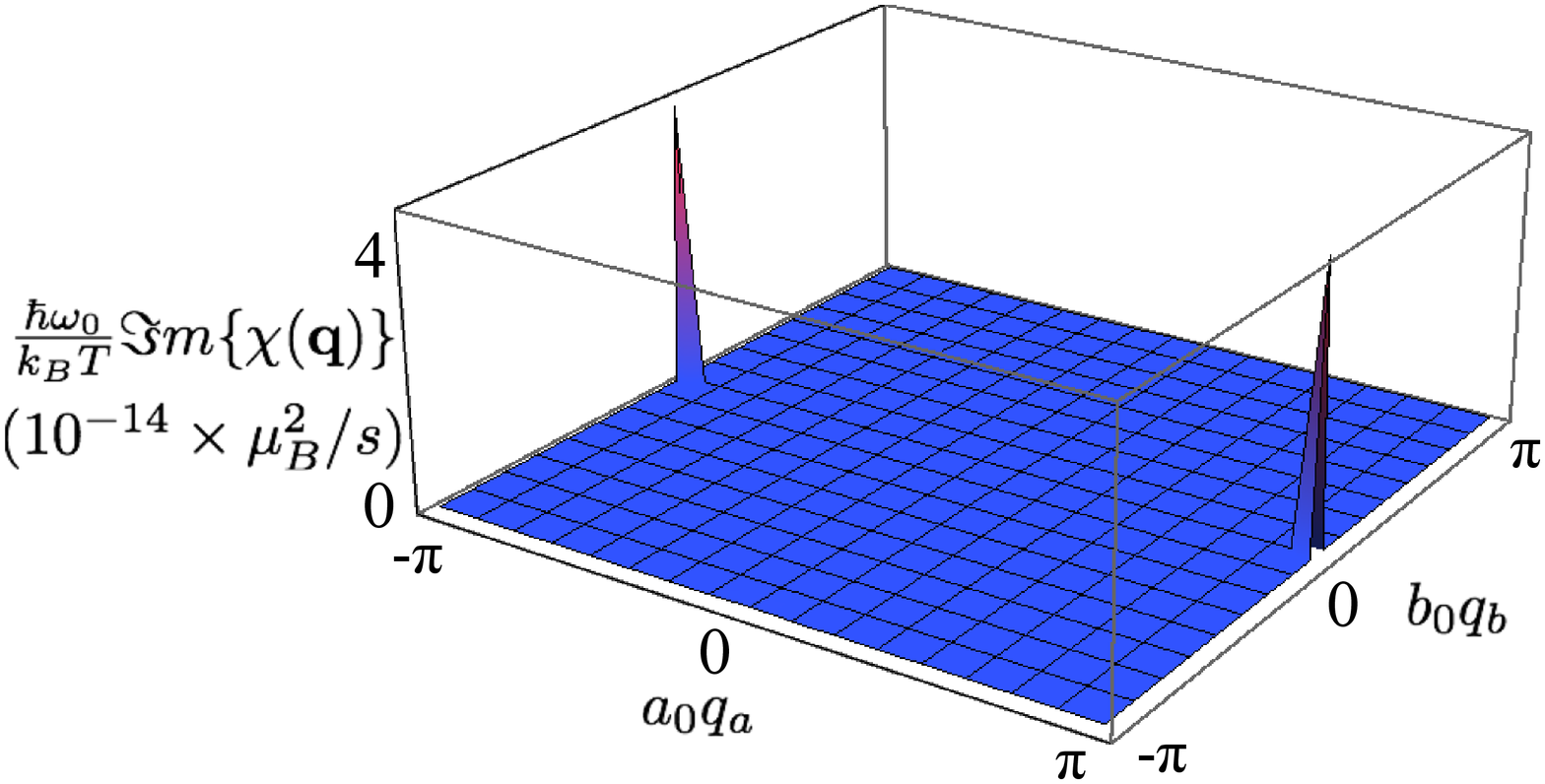}}
\subfigure[]{\includegraphics[width=0.45\textwidth]{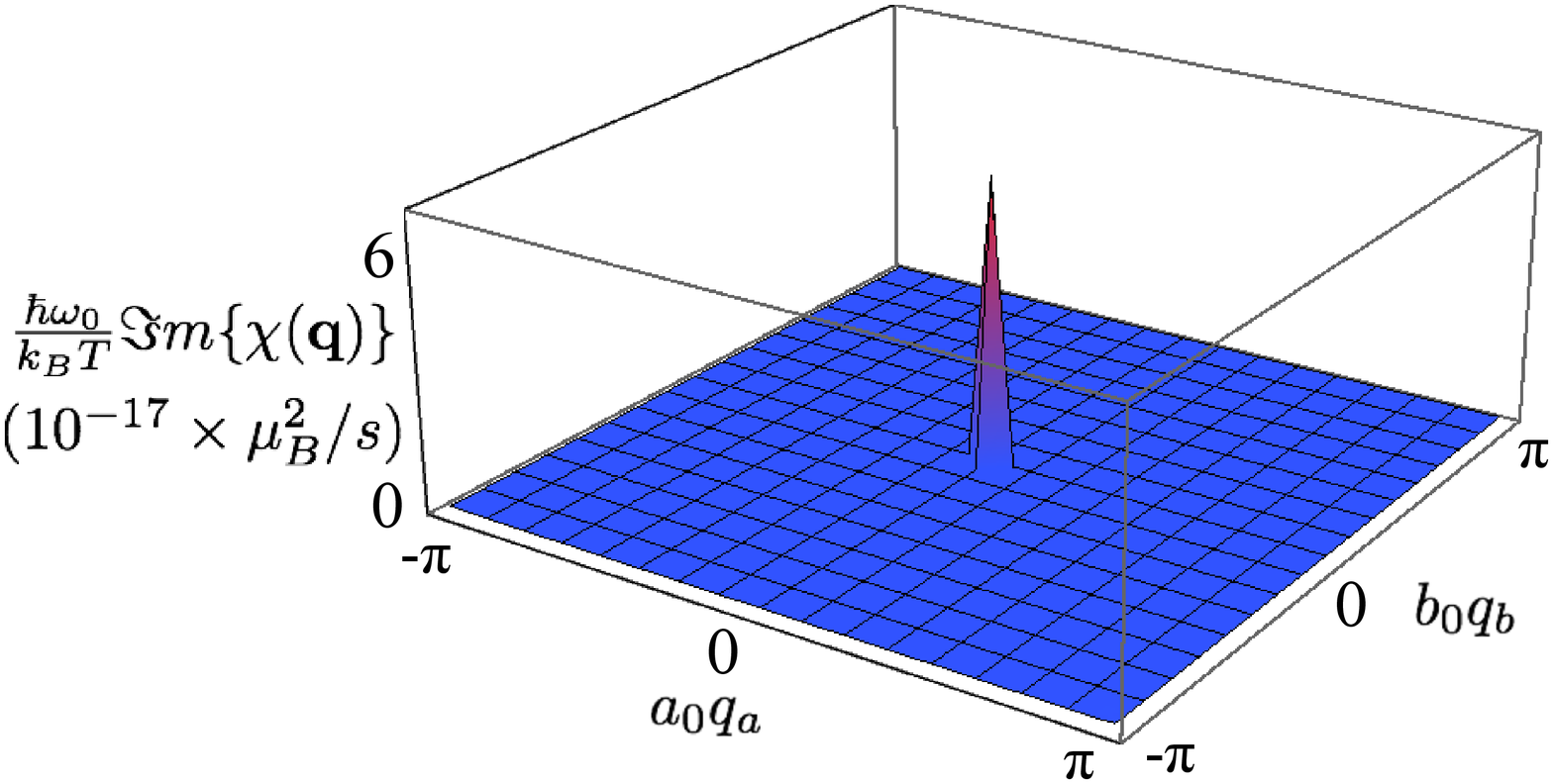}}
\caption{\footnotesize{(Color online). The ${\bf q}$ dependence of the imaginary part of the longitudinal, dynamic susceptibility at a) $q_c=\pi/c_0$ and b) $q_c=0$, as predicted by Eq.~(\ref{eq:staggeredfieldsuscep}) and Eq.~(\ref{eq:constfieldsuscep}). The peak at \mbox{${\bf q}=(\pi/a_0,0,q_c)$} is approximately 1000 times larger than that at ${\bf q}=0$. We use the parameters \mbox{$v_a=v_b=280 \ meV \AA$}, \mbox{$v_c=57 \ meV \AA$} and \mbox{$\Delta = 9.8 \ meV$} from [\onlinecite{matan09}] and \mbox{$\chi_\perp=1.2\times 10^{-3}$} from [\onlinecite{ning09}]. The temperature is \mbox{$k_BT = 1 \ meV$}.
 }}
\label{fig:susceptibility}
\end{figure}

The imaginary part of the susceptibility has a large peak at the ordering vector ${\bf q}={\bf Q}=(\pi/a_0,0,\pi/c_0)$ and a smaller peak at ${\bf q}=0$, as shown in Fig.~(\ref{fig:susceptibility}). The width of these peaks is controlled by the temperature, $T$, and, for all other wavevectors, the susceptibility is exponentially suppressed. For a realistic set of parameters the peak at ${\bf q}={\bf Q}$ is three of orders of magnitude larger than that at ${\bf q}=0$, as illustrated in Fig.~(\ref{fig:susceptibility}).


\subsection{The relaxation rate with field in the $[110]$-direction}
\label{subsec:abfield}


We are now in a position to determine the relaxation rate for an external field applied in the $[110]$-direction. The form factor is non-zero at \mbox{${\bf q}={\bf Q}=(\pi/a_0,0,\pi/c_0)$}, which is the peak in the imaginary part of the susceptibility. Thus the neighbourhood of this point in momentum space will dominate the integral for the relaxation rate.


Expanding the form factor in Eq.~(\ref{eq:abformfactorallq}) around ${\bf q}={\bf Q}$ leads to,
 \begin{align}
\mathcal{F}({\bf q}\approx {\bf Q}, h^{110}_{\sf ext}) &\approx 16 (\mathcal{A}^{ca})^2. 
\end{align}
Substituting this into Eq.~(\ref{eq:T1general1}), along with Eq.~(\ref{eq:staggeredfieldsuscep}), and making the coordinate transformation,
\begin{align}
k_{1\alpha}=&v_\alpha k_\alpha  \nonumber  \\
 k_{2\alpha}=&v_\alpha(k_\alpha+q_\alpha-Q_\alpha),
 \label{eq:coordtransst}
\end{align}
gives,
\begin{align}
&\frac{1}{T_1(h^{110}_{\sf ext})} \approx
\frac{4 \pi \hbar m_0^2  (\mathcal{A}^{ca})^2\gamma_N^2 (a_0b_0c_0)^2}{\chi_\perp^2 \bar{v}_s^6}   \nonumber \\
&\times \int_{cone} \frac{d^3k_1}{(2\pi)^3}  \int_{cone} \frac{d^3k_2}{(2\pi)^3} 
 \frac{1}{(\hbar\omega_{{\bf k}_1})^2}
n_B( \omega_{{\bf k}_1})(n_B( \omega_{{\bf k}_1})+1) \nonumber \\
&  \qquad  \qquad \qquad \times \delta(\hbar\omega_{{\bf k}_1}-\hbar\omega_{{\bf k}_2}),
\end{align}
where $m_0=g_l \mu_B S$, $\omega_{{\bf k}}=\omega_{1,{\bf k}}$,
\begin{align}
 \bar{v}_s=(v_av_bv_c)^{\frac{1}{3}},
\end{align}
is the geometric mean of the spin wave velocities and the integrals are over a cone of spin wave excitations.


The density of states, which is given in Appendix~\ref{App:dos}, can be used to transform the integral over momentum into one over energy, resulting in,
\begin{align}
\frac{1}{T_1(h^{110}_{\sf ext})} \approx& 
\frac{ \hbar m_0^2  \gamma_N^2  (a_0b_0c_0)^2}{\pi^3 \bar{v}_s^6 \chi_\perp^2} (\mathcal{A}^{ca})^2  \nonumber \\
& \times \int_{\Delta}^\infty d\epsilon  (\epsilon^2-\Delta^2)\frac{e^{\epsilon/k_BT}}{\left( e^{\epsilon/k_BT} -1\right)^2}.
 \end{align}
The required energy integrals are evaluated in Appendix~\ref{App:boseintegrals}, and it follows that the relaxation rate is,
 \begin{align}
\frac{1}{T_1(h^{110}_{\sf ext})} \approx 
 (\mathcal{A}^{ca})^2  \ C_{st,1} \
 \Phi_{st,1}  \left( \frac{k_B T}{\Delta} \right),
 \label{eq:T1hab}
\end{align} 
where,
\begin{eqnarray} 
C_{st,1}=\frac{ 2\hbar m_0^2  \gamma_N^2  (a_0b_0c_0)^2\Delta^3}{\pi^3 \bar{v}_s^6 \chi_\perp^2},
\label{eq:Cst1}
\end{eqnarray} 
and,
\begin{eqnarray} 
\Phi_{st,1} (x) = x^2 \mathrm{Li}_1(e^{-1/x}) + x^3 \mathrm{Li}_2(e^{-1/x}),
\end{eqnarray} 
with
\mbox{$\mathrm{Li}_m(z) = \sum_{l=0}^\infty z^l/l^m$} the m$^{th}$ polylogarithm of $z$. 


\subsection{The relaxation rate with field in the $c$-direction}
\label{subsec:cfield}


We now turn to the relaxation rate with the external field applied in the $c$-direction. The form factor is qualitatively different from that with the field applied in the $[110]$-direction, since it is no longer peaked at the ordering vector but is in fact zero at this point. Expanding the form factor to lowest order around the wavevectors ${\bf q}= {\bf Q}$ and ${\bf q}= 0$ gives,
 \begin{align}
\mathcal{F}({\bf q}\approx {\bf Q}, h^{c}_{\sf ext}) &\approx 4 (\mathcal{A}^{aa})^2a_0^2 (q_a-Q_a)^2  \nonumber \\
&\qquad +4 (\mathcal{A}^{ba})^2 b_0^2(q_b-Q_b)^2  \nonumber \\
\mathcal{F}({\bf q}\approx 0, h^{c}_{\sf ext}) &\approx 16(\mathcal{A}^{aa})^2.
\end{align}


There are thus two main contributions to the relaxation rate. The first, from the region around  ${\bf q}=0$, we denote as $1/T_1^{un}$ and use Eq.~(\ref{eq:constfieldsuscep}) to write it as,
\begin{align}
&\frac{1}{T^{un}_1(h^c_{\sf ext})} \approx  (\mathcal{A}^{aa})^2 
 \frac{16 \pi \hbar (g_l\mu_B)^2 \gamma_N^2   (a_0b_0c_0)^2}{\bar{v}_s^6}
 \nonumber \\
&\times \int_{cone} \frac{d^3k_1}{(2\pi)^3}  \int_{cone} \frac{d^3k_2}{(2\pi)^3} 
 \frac{1}{(\hbar\omega_{{\bf k}_1})^2}
n_B( \omega_{{\bf k}_1})(n_B( \omega_{{\bf k}_1})+1) \nonumber \\
&  \qquad  \qquad \qquad \times \delta(\hbar\omega_{{\bf k}_1}-\hbar\omega_{{\bf k}_2}).
\end{align}
where the coordinate transformation,
\begin{align}
k_{1\alpha}=&v_\alpha k_\alpha  \nonumber  \\
 k_{2\alpha}=&v_\alpha(k_\alpha+q_\alpha),
\end{align}
has been applied.
Making use of the results in Appendix~\ref{App:dos} and Appendix~\ref{App:boseintegrals}, leads to,
 \begin{align}
\frac{1}{T^{un}_1(h^c_{\sf ext})} \approx & (\mathcal{A}^{aa})^2
C_{un,1} \
\Phi_{un,1} \left( \frac{k_B T}{\Delta}  \right),
\end{align}
where,
 \begin{align}
C_{un,1}=\frac{8 (g\mu_B)^2  \hbar  \gamma_N^2   (a_0b_0c_0)^2 \Delta^5}{\pi^3\bar{v}_s^6} 
\label{eq:Cun1},
\end{align}
and,
\begin{align} 
\Phi_{un,1} (x) =& \ x^2 \mathrm{Li}_1(e^{-1/x}) + 5x^3 \mathrm{Li}_2(e^{-1/x}) \nonumber \\
&+12x^4 \mathrm{Li}_3(e^{-1/x})+12x^5 \mathrm{Li}_4(e^{-1/x}).
\end{align} 


The second contribution to the relaxation rate is from the region surrounding ${\bf q}={\bf Q}$. This is suppressed relative to the $[110]$ field direction by the vanishing of the form factor at this point. We denote this contribution as $1/T_1^{st}$ and find,
\begin{align}
&\frac{1}{T^{st}_1(h^c_{\sf ext})} \approx
\frac{ \pi \hbar m_0^2 \gamma_N^2   (a_0b_0c_0)^2 }{\bar{v}_s^6 \chi_\perp^2}
\int_{cone} \frac{d^3k_1}{(2\pi)^3}  \int_{cone} \frac{d^3k_2}{(2\pi)^3} \nonumber \\
 & \frac{1}{2}\left[  (\mathcal{A}^{aa})^2 \left( \frac{a_0}{v_a} \right)^2 \left(  k_{2a}^2+ k_{1a}^2 \right)  
 + (\mathcal{A}^{ba})^2  \left( \frac{b_0}{v_b} \right)^2 \left(  k_{2b}^2+ k_{1b}^2 \right) \right] \nonumber \\
 & \frac{n_B( \omega_{{\bf k}_1})(n_B( \omega_{{\bf k}_1})+1)}{(\hbar\omega_{{\bf k}_1})^2}
\ \delta(\hbar\omega_{{\bf k}_1}-\hbar\omega_{{\bf k}_2}),
\label{eq:T1ksq}
\end{align}
where the coordinate transformation defined in Eq.~(\ref{eq:coordtransst}) has been applied.
Making use of the spectral functions derived in Appendix~\ref{App:dos} and the integrals evaluated in Appendix~\ref{App:boseintegrals} leads to a  relaxation rate,
\begin{align}
&\frac{1}{T^{st}_1(h^c_{\sf ext})} \approx C_{st,k^2}
 \left[   \frac{a_0^2(\mathcal{A}^{aa})^2}{v_a^2} 
  +   \frac{b_0^2(\mathcal{A}^{ba})^2}{v_b^2}  \right]
\Phi^{3d}_{st,k^2} \left(\frac{k_B T}{\Delta} \right)
\end{align}
where,
\begin{align}
C_{st,k^2} =\frac{4 \hbar  m_0^2 \gamma_N^2  (a_0b_0c_0)^2\Delta^5}{3\pi^3 \chi_\perp^2 \bar{v}_s^6},
\label{eq:Cstksq}
\end{align}
and,
\begin{align} 
\Phi_{st,k^2} (x) =  &x^3 \mathrm{Li}_2(e^{-1/x}) 
+3x^4 \mathrm{Li}_3(e^{-1/x}) \nonumber \\
&+3x^5 \mathrm{Li}_4(e^{-1/x}).
\end{align} 


The total relaxation rate for external field in the $c$-direction is given by the sum of the two contributions,
\begin{align}
\frac{1}{T_1(h^c_{\sf ext})} \approx \frac{1}{T^{un}_1(h^c_{\sf ext})}+\frac{1}{T^{st}_1(h^c_{\sf ext})}.
\label{eq:T1hc}
\end{align}


\subsection{Comparison with experiment}


Having derived theoretical predictions for the relaxation rate, we compare to experimental data, 
and show that the two are consistent at a quantitative level.


The experimental data for BaFe$_2$As$_2$ are separable into an isotropic term, which is linear in 
temperature, and an activated anisotropic term, which has a more complicated temperature dependence. 
The isotropic contribution to the relaxation rate is likely due to a fluid of conduction electrons associated 
with ungapped portions of the Fermi surface.   
The linear temperature dependence would then be attributable to a 
Korringa-type relaxation rate\cite{slichter,moriya63}. 


We have previously argued that, at low temperatures, the interaction between these two electron fluids is 
negligible\cite{smerald10}. However, the origin of this isotropic relaxation rate is not pertinent to this paper. 
We consider that the anisotropic term comes from the scattering of thermally excited spin waves and model 
the relaxation rate using the results of Sections~\ref{subsec:abfield}~and~\ref{subsec:cfield}.

\begin{figure}[h]
\centering
\includegraphics[width=0.45\textwidth]{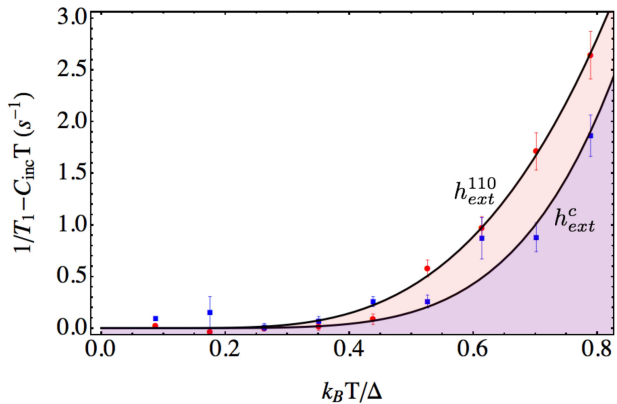}
\caption{\footnotesize{(Color online). Fits to the NMR relaxation rate data shown in the inset of 
Fig.~(\ref{fig:Ba122inequivAs}).  
For external field in the $[110]$-direction (red circles) we fit Eq.~(\ref{eq:T1hab}), while for external field parallel 
to the $c$-axis (blue squares) we fit Eq.~(\ref{eq:T1hc}).  
The same, linear $T$, isotropic term has been fitted and subtracted from both data sets.  
In fitting the anisotropic contribution to $1/T_1$ we allow a single free parameter for each external 
field orientation: for the $[110]$-direction $(\mathcal{A}^{ca})^2  C_{st,1}$; and for the $c$-direction 
$(\mathcal{A}^{aa})^2 C_{un,1}$.  
This provides convincing fits to the data, and we show in the text that estimating these fit parameters 
from independent experiments leads to quantitative agreement.  
We use $\Delta=9.8 \ meV$, taken from neutron scattering experiments\cite{matan09}.
}}
\label{fig:Ba122T1}
\end{figure}

Describing the low temperature region of the experimental data with a function $1/T_1\approx C_{inc} T$, 
results in good fits to both data sets with  $C_{inc}=0.032 s^{-1}K^{-1}$.  
In Fig.~(\ref{fig:Ba122T1})  this isotropic term is subtracted from the data and we concentrate 
on fitting the anisotropic contribution to the relaxation rate using the theory described above.  


For external field in the $[110]$-direction we treat $(\mathcal{A}^{ca})^2C_{st,1}$ as a free parameter.  
This results in a convincing fit to the data for,
\begin{align}
(\mathcal{A}^{ca})^2  C_{st,1}  \approx 7.5\ s^{-1}.
\label{eq:abfitparameter}
\end{align}
Eq.~(\ref{eq:Cst1}) gives $C_{st,1}$ in terms of parameters that have been measured in independent 
experiments. 
Substituting in the values in Table~\ref{tab:Ba122experimental} implies,
\begin{align}
0.13 < (\mathcal{A}^{ca})^2  C_{st,1} <31 \ s^{-1},
\end{align}
where the error is predominantly due to the uncertainty in the spin-wave velocities.  
Therefore we find that the fits to the NMR data are in {\it quantitative} agreement with 
independent experiments, within the limits set by experimental error on measurements 
of the input parameters of the theory.


We have previously shown that Eq.~(\ref{eq:T1hab}) also provides convincing fits to NMR $1/T_1$ 
relaxation rate data for SrFe$_2$As$_2$, with magnetic field parallel to $[110]$ [\onlinecite{smerald10}].   
Similarly, Klanjsek {\it et al}. have found good agreement with $1/T_1$ data for NaFeAs [\onlinecite{klanjsek11}].


\begin{table}
\begin{center}
\footnotesize
  \begin{tabular}{| c | c | c | c |}
    \hline
Quantity    & Value  & Method & Ref  \\ \hline 
$\Delta$ & 9.8(4) meV & Neutrons  & [\onlinecite{matan09}]  \\ \hline
 $ \chi_\perp$ & $10^{-4}$ emu/mol  $=1.2\times 10^{-3}$ meV$^{-1}$ &Knight shift &  [\onlinecite{ning09}]  \\ \hline
$[a_0,b_0,c_0]$ & $[2.80,2.79,6.47] \ \AA$ & Neutrons & [\onlinecite{huang08}]  \\ \hline
$m_0$ & 0.87 $\mu_B$ & Neutrons & [\onlinecite{huang08}]  \\ \hline
$\mathcal{A}^{ca}$ & 0.43 T/$\mu_B$  & NMR & [\onlinecite{kitagawa08}]  \\ \hline
$\mathcal{A}^{aa}$ & 0.66 T/$\mu_B$  & Knight shift & [\onlinecite{kitagawa08}]  \\ \hline
$\gamma_N^{As}$ & 4.6$\times 10^7$ T$^{-1}$s$^{-1}$   & Tabulated &  \\ \hline
$v_a$ &  280(150) meV$\AA$  & Neutrons  &  [\onlinecite{matan09}] \\ \hline
$v_b$ &  280(150) meV$\AA$  & Neutrons &  [\onlinecite{matan09}] \\ \hline
$v_c$ &  57(7) meV$\AA$  & Neutrons &  [\onlinecite{matan09}] \\ \hline
$|{\bf h}_{\sf ext}|$ & 1.5T & NMR  & [\onlinecite{kitagawa08}]  \\ \hline
    \end{tabular}
\end{center} 
\caption{\footnotesize{Parameters used to fit  the $1/T_1$ relaxation rate in BaFe$_2$As$_2$. 
}}
\label{tab:Ba122experimental}
\end{table}


For external field in the $c$-direction we fix the ratio,
\begin{align}
C_{st,k^2}= \frac{m_0^2 a_0^2}{12\chi_\perp^2v_a^2} C_{un,1} ,
\end{align}
and in the absence of other information assume \mbox{$\mathcal{A}^{aa}\approx\mathcal{A}^{ba}$}.  
As shown in Fig.~(\ref{fig:Ba122T1}), fitting Eq.~(\ref{eq:T1hc}) to the data with the single free 
parameter $ (\mathcal{A}^{aa})^2 C_{un,1}$ gives a convincing fit to the data for,
\begin{align}
  0.12< (\mathcal{A}^{aa})^2 C_{un,1} <0.5  \ s^{-1},
  \label{eq:cfitparameter}
\end{align}
where the uncertainty comes from the spin wave-velocities $v_a$ and $v_b$.
Estimating $C_{un,1}$ using the values in Table~\ref{tab:Ba122experimental} gives,
\begin{align}
 0.002 < (\mathcal{A}^{aa})^2 C_{un,1} < 0.25 \ s^{-1}.
\end{align}
Again, within the bounds of experimental error, we find quantitative agreement between 
theory and experiment. 


In summary, the theory for $1/T_1$ developed throughout this section is in {\it quantitative} 
agreement with the experimental data of Kitagawa {\it et al}\cite{kitagawa08}. 
This demonstrates the importance of taking angular resolution into account when calculating relaxation rates. 
We will now go on to generalise these results for an arbitrary strength and direction of the external magnetic field. 


\section{Extension of theory to arbitrary orientation and magnitude of applied field}
\label{sec:arbitraryfield}


In this section we develop a theory of the relaxation rate for arbitrary orientation and magnitude of the 
external magnetic field.
This follows from determining the general expression for the form factor in BaFe$_2$As$_2$, and then 
combining this with the above calculation of the longitudinal susceptibility.


In order to find the form factor for an arbitrary magnitude and orientation of magnetic field, it is necessary to 
study the rotation matrices that transform between the $(x,y,z)$ coordinates [those in which $z$ is aligned 
with ${\bf h}_{\sf nuc}$] and the $(a,b,c)$ coordinates [those aligned with the crystal axes].  
Consider a rotation matrix $ \underline{\underline{R}}_{{\bf h}_{\sf ext}}$ that rotates a vector from 
the $(a,b,c)$ coordinate system into the $(x,y,z)$ coordinate system.  
The action of this matrix on the objects of interest is,
 \begin{align}
m_{{\bf q}}^\xi &=\sum_\alpha R^{\xi\alpha}_{{\bf h}_{\sf ext}}m_{{\bf q}}^\alpha \nonumber \\
 \mathcal{A}_i^{\xi\psi}  &= \sum_{\alpha,\beta}  R^{\xi\alpha}_{{\bf h}_{\sf ext}} \mathcal{A}_i^{\alpha\beta} \left(R^{-1}_{{\bf h}_{\sf ext}}\right)^{\beta\psi} \nonumber \\
 &= \sum_{\alpha,\beta}  R^{\xi\alpha}_{{\bf h}_{\sf ext}} R^{\psi\beta}_{{\bf h}_{\sf ext}} \mathcal{A}_i^{\alpha\beta}   \nonumber \\
\chi^{\xi\psi}({\bf q},\omega_0) &=  \sum_{\alpha,\beta}  R^{\xi\alpha}_{{\bf h}_{\sf ext}} R^{\psi\beta}_{{\bf h}_{\sf ext}}  \chi^{\alpha\beta}({\bf q},\omega_0),
 \label{eq:coordrotation}
\end{align}
where $\alpha, \beta=\{ a,b,c \}$ and $\xi, \psi=\{ x,y,z \}$. 


These rotation matrices can be used to transform Eq.~(\ref{eq:T1suscep}) for the relaxation rate into,
\begin{align}
&\frac{1}{T_1({\bf h}_{\sf ext})} =  \lim_{\omega_0\rightarrow 0}\frac{\gamma_N^2}{2N}  k_BT \sum_{{\bf q},\alpha,\beta,\gamma,\delta}  \nonumber \\
&
\left[ R^{x \gamma}_{{\bf h}_{\sf ext}} R^{x \delta}_{{\bf h}_{\sf ext}} + R^{y \gamma}_{{\bf h}_{\sf ext}} R^{y \delta}_{{\bf h}_{\sf ext}} \right] 
\mathcal{A}_{\bf q}^{\gamma \alpha} \mathcal{A}_{-{\bf q}}^{\delta \beta}
\frac{ \Im m \left\{ \chi^{\alpha \beta}({\bf q},\omega_0) \right\} }{\hbar \omega_0},
\label{eq:T1abc}
\end{align}
where $\gamma, \delta=\{ a,b,c \}$. 


Since only the longitudinal susceptibility, \mbox{$\chi^{aa}({\bf q},\omega_0)=\chi_\parallel({\bf q},\omega_0)$}, is relevant to the relaxation process in BaFe$_2$As$_2$, it follows that, 
 \begin{align}
\frac{1}{T_1({\bf h}_{\sf ext})} &=  \lim_{\omega_0\rightarrow 0}\frac{\gamma_N^2}{2N}  k_BT \hspace{-3mm} \sum_{{\bf q}\in \mathrm{PMBZ}}  \hspace{-3mm}\mathcal{F}({\bf q},{\bf h}_{\sf ext})
\frac{ \Im m \left\{ \chi_{\parallel}({\bf q},\omega_0) \right\} }{\hbar \omega_0},
\label{eqapp:T1general1}
\end{align}
where  the form factor that couples to the longitudinal spin fluctuations is,
\begin{align}
\mathcal{F}({\bf q},{\bf h}_{\sf ext})
&= \sum_{\gamma,\delta} 
 \left[ R^{x \gamma}_{{\bf h}_{\sf ext}} R^{x \delta}_{{\bf h}_{\sf ext}} + R^{y \gamma}_{{\bf h}_{\sf ext}} R^{y \delta}_{{\bf h}_{\sf ext}} \right] 
\mathcal{A}_{\bf q}^{\gamma a} \mathcal{A}_{-{\bf q}}^{\delta a}.
\label{eq:formfactorgeneral}
\end{align}

\begin{figure}[h]
\centering
\includegraphics[width=0.36\textwidth]{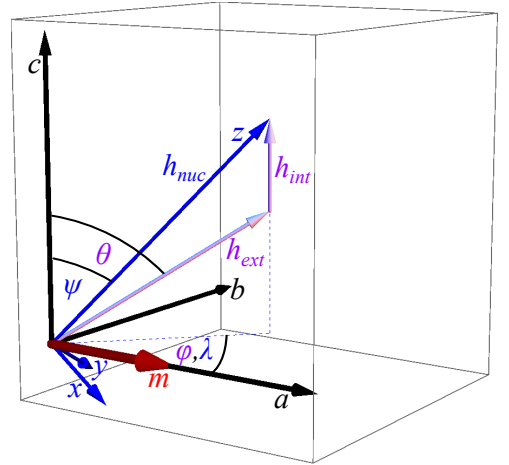}
\caption{\footnotesize{(Color online). 
The relationship between the coordinate system of the crystal axes of 
BaFe$_2$As$_2$, $(a,b,c)$, 
and the coordinate system of the effective magnetic field 
${\bf h}_{\sf nuc}$ at the $^{75}$As nucleus, $(x,y,z)$. 
In the magnetically ordered phase, the electron moments, ${\bf m}$, are orientated along the 
crystallographic $a$-axis.
The interaction between the $^{75}$As nucleus and these magnetically ordered electrons 
gives rise to an effective internal magnetic field, ${\bf h}_{\sf int}$, directed along the $c$-axis.
The total field \mbox{${\bf h}_{\sf nuc}={\bf h}_{\sf ext}+{\bf h}_{\sf int}$} is the sum of ${\bf h}_{\sf int}$ 
and the external magnetic field, ${\bf h}_{\sf ext}$, applied during NMR experiments.
The orientation of the external field, ${\bf h}_{\sf ext}$, relative to the crystal axes 
[shown here with polar angles $(\theta,\phi)$] can be varied at will by rotating the 
sample in a goniometer.
This in turn changes the orientation of the total effective field ${\bf h}_{\sf nuc}$ 
[shown here with polar angles $(\psi,\lambda)$].  
}}
\label{fig:coordinates}
\end{figure}

For an external field of arbitrary magnitude, the assumption $ {\bf h}_{\sf nuc}\approx {\bf h}_{\sf ext}$ is no longer valid. As such it is natural to define two sets of angles, as shown in Fig.~(\ref{fig:coordinates}). The first set of angles, $(\psi,\lambda)$, describe the orientation of $ {\bf h}_{\sf nuc}$ in the $(a,b,c)$ crystallographic coordinate system.  $\psi$ is the angle between the $z$-axis and the $c$-axis and $\lambda$ is the angle between the projection of the $z$-axis onto the $ab$ plane and the $a$-axis. The rotation matrix is thus,
\begin{align}
&\underline{\underline{\mathcal{R}}}= \nonumber \\
&\left( \hspace{-1mm}
\begin{array}{ccc}
\sin^2\lambda+\cos\psi\cos^2\lambda & -\sin 2\lambda\sin^2\frac{\psi}{2} & \cos\lambda \sin\psi \\
-\sin 2\lambda\sin^2\frac{\psi}{2} & \cos^2\lambda+\cos\psi\sin^2\lambda & \sin\lambda \sin\psi \\
-\cos\lambda \sin\psi & -\sin\lambda \sin\psi &  \cos\psi
\end{array}
\hspace{-1mm} \right).
\label{eq:rotationmatrix}
\end{align}
The second set of angles, $(\theta,\phi)$, describe the orientation of  $ {\bf h}_{\sf ext}$ in the $(a,b,c)$ crystallographic coordinate system. These are the experimentally accessible set of angles, where $\theta$ is the angle between the $c$-axis and ${\bf h}_{\sf ext}$, and $\phi$ is the angle between the $a$-axis and the projection of ${\bf h}_{\sf ext}$ onto the $ab$-plane.


For the As nucleus in BaFe$_2$As$_2$, Eq.~(\ref{eq:hintc}) gives,
\mbox{$\langle {\bf h}_{\sf int} \rangle=\left(0,0,\pm |\langle {\bf h}_{\sf int} \rangle| \right)$}, and it follows that,
\begin{align}
\lambda &=\phi, \nonumber \\ 
\tan\psi &= \frac{|{\bf h}_{\sf ext}|\sin\theta }{|{\bf h}_{\sf ext}| \cos\theta  \pm | \langle{\bf h}_{\sf int}\rangle |}.
\end{align}
Thus the $(\psi,\lambda)$ angles that enter the theory can be expressed in terms of the known angles $(\theta,\phi)$. In the high external field regime, $|{\bf h}_{\sf ext}|\gg |\langle {\bf h}_{\sf int} \rangle|$, and therefore $\psi=\theta$ and $\lambda=\phi$.


The form factor that follows from substituting Eq.~(\ref{eq:rotationmatrix}) into Eq.~(\ref{eq:formfactorgeneral}) is,
\begin{align}
\mathcal{F}({\bf q},{\bf h}_{\sf ext})
=&  
\left( \cos^2\psi+\sin^2\lambda \sin^2\psi \right) \mathcal{A}_{\bf q}^{aa} \mathcal{A}_{-{\bf q}}^{aa} \nonumber \\
&+ \left( \cos^2\psi+\cos^2\lambda \sin^2\psi \right) \mathcal{A}_{\bf q}^{ba} \mathcal{A}_{-{\bf q}}^{ba} \nonumber \\
&+ \sin^2\psi \ \mathcal{A}_{\bf q}^{ca} \mathcal{A}_{-{\bf q}}^{ca} \nonumber \\
&- \frac{1}{2}\sin2\lambda \sin^2\psi  \left( \mathcal{A}_{\bf q}^{aa} \mathcal{A}_{-{\bf q}}^{ba} +\mathcal{A}_{\bf q}^{ba} \mathcal{A}_{-{\bf q}}^{aa}  \right) \nonumber \\
&+ \frac{1}{2}\cos\lambda \sin2\psi  \left( \mathcal{A}_{\bf q}^{aa} \mathcal{A}_{-{\bf q}}^{ca} +\mathcal{A}_{\bf q}^{ca} \mathcal{A}_{-{\bf q}}^{aa}  \right) \nonumber \\
&+ \frac{1}{2}\sin\lambda \sin2\psi  \left( \mathcal{A}_{\bf q}^{ba} \mathcal{A}_{-{\bf q}}^{ca} +\mathcal{A}_{\bf q}^{ca} \mathcal{A}_{-{\bf q}}^{ba}  \right).
\end{align}
Eq.~(\ref{eq:Aqmatrix}) can be used to re-express this as,
\begin{align}
\mathcal{F}({\bf q},{\bf h}_{\sf ext})&=
16 \left( \cos^2\psi +\sin^2\lambda  \sin^2\psi \right) (\mathcal{A}^{aa}c_a  c_b)^2 \nonumber \\
&+16 \left( \cos^2\psi +\cos^2\lambda  \sin^2\psi \right) (\mathcal{A}^{ba} s_a s_b)^2 \nonumber \\
& +16 \sin^2\psi (\mathcal{A}^{ca}  s_a c_b)^2   \nonumber \\
& +16\sin2\lambda \sin^2\psi \ \mathcal{A}^{aa}\mathcal{A}^{ba} c_ac_bs_as_b,
\label{eq:Ba122formfactor}
\end{align}
where $c_a$, $c_b$, $s_a$ and $s_b$ are given in Eq.~(\ref{eq:casa}). Approximating the form factor close to ${\bf q}={\bf Q}$ gives,
\begin{align}
& \mathcal{F}({\bf q}\approx{\bf Q},{\bf h} _{\sf ext}) \approx   \nonumber \\
& \qquad  16 (\mathcal{A}^{ca})^2  \sin^2\psi \nonumber \\
& \qquad   +4(\mathcal{A}^{aa})^2 \left( \cos^2\psi +\sin^2\lambda  \sin^2\psi \right) a_0^2(q_a-Q_a)^2  \nonumber \\
& \qquad  +4(\mathcal{A}^{ba})^2 \left( \cos^2\psi +\cos^2\lambda  \sin^2\psi \right) b_0^2(q_b-Q_b)^2  \nonumber \\
& \qquad   +4 \mathcal{A}^{aa}\mathcal{A}^{ba} \sin2\lambda \sin^2\psi \ a_0b_0(q_a-Q_a)(q_b-Q_b) ,
\end{align}
while close to ${\bf q}=0$ the leading contribution is,
\begin{align}
& \mathcal{F}({\bf q}\approx 0,{\bf h} _{\sf ext}) \approx  
16(\mathcal{A}^{aa})^2 \left( \cos^2\psi +\sin^2\lambda  \sin^2\psi \right).
\end{align}

\begin{figure}[h]
\centering
\includegraphics[width=0.45\textwidth]{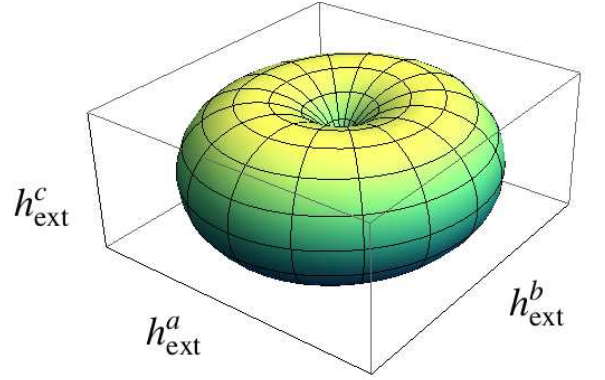}
\caption{\footnotesize{(Color online). 
``Doughnut'' shaped prediction for the variation of the $^{75}$As NMR $1/T_1$ 
relaxation rate of BaFe$_2$As$_2$ with the orientation of the external magnetic field 
${\bf h}_{\sf ext}$, at fixed temperature.
The value of $1/T_1$ is represented by the radial distance of the surface from the origin, 
and calculated from Eq.~(\ref{eq:T1angleresolved}) using the parameters given 
in Table~\ref{tab:Ba122experimental}, for $T=0.6\Delta$, $|{\bf h}_{\sf ext}|=6T$.  
The relaxation rate is a minimum when the external field is applied along the $c$-axis, 
parallel to the internal field $\langle {\bf h}_{\sf int} \rangle$.
 }}
\label{fig:Ba122doughnut}
\end{figure}

The general form of the relaxation rate is now accessible.   
The techniques outlined in Section~\ref{sec:datafit} can be used to find, 
\begin{align}
&\frac{1}{T_1({ \bf h}_{\sf ext})} \approx  (\mathcal{A}^{ca})^2 \sin^2\psi \ C_{st,1}\
 \Phi_{st,1}  \left( \frac{k_B T}{\Delta} \right) \nonumber \\
 & \quad +(\mathcal{A}^{aa})^2
\left( \cos^2\psi +\sin^2\lambda  \sin^2\psi \right)  C_{un,1} \
\Phi_{un,1} \left( \frac{k_B T}{\Delta}  \right) \nonumber \\
&\quad + C_{st,k^2} 
 \left[  \left( \cos^2\psi +\sin^2\lambda  \sin^2\psi \right) \frac{a_0^2(\mathcal{A}^{aa})^2}{v_a^2} \right. \nonumber \\
&\quad  \left. +   \left( \cos^2\psi +\cos^2\lambda  \sin^2\psi \right)\frac{b_0^2(\mathcal{A}^{ba})^2}{v_b^2}  \right]
\Phi_{st,k^2} \left(\frac{k_B T}{\Delta} \right).
\label{eq:T1angleresolved}
\end{align}
This equation for $1/T_1$ leads to a ``doughnut''-shaped angular dependence, as illustrated 
in Fig.~(\ref{fig:Ba122doughnut}).
The rate is largest when ${\bf h}_{\sf ext}$ is orientated in the $ab$-plane, and smallest for 
${\bf h}_{\sf ext}$ in the $c$-direction. 
There is a small difference between external field aligned in the $a$-direction and in the 
$b$-direction, with the $b$-direction being faster.


\begin{figure}[h]
\centering
\includegraphics[width=0.45\textwidth]{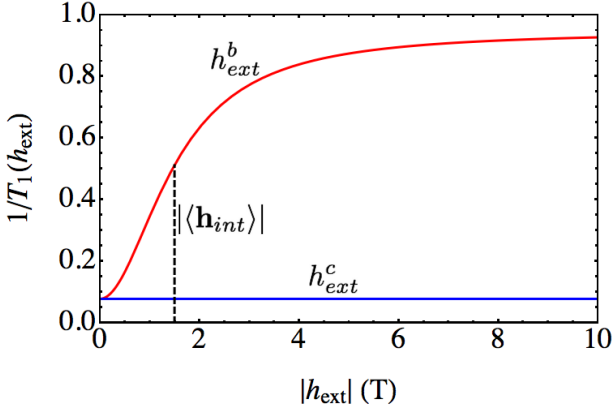}
\caption{\footnotesize{(Color online). Theoretical predictions for the anisotropic contribution to the relaxation rate as the magnitude of the external field, ${ \bf h}_{\sf ext}$, is varied. For an external field applied in the $c$-direction there is no dependence on magnitude. For an external field applied in the $b$-direction the relaxation rate is strongly dependent on magnitude, as described by Eq.~(\ref{eq:Ba122hextvariationb}). We set $T=0.6\Delta$ and use the parameters obtained from fits to BaFe$_2$As$_2$ data (Eq.~(\ref{eq:abfitparameter}) and Eq.~(\ref{eq:cfitparameter})).
}}
\label{fig:Ba122hextmagnitude}
\end{figure}

As shown in Fig.~(\ref{fig:Ba122hextmagnitude}), the relaxation rate depends not only on the orientation of the external magnetic field, but also on the magnitude. 
This arises from the fact that it is the orientation of ${\bf h}_{\sf nuc}$ that determines the relaxation rate. 
When ${\bf h}_{\sf ext}=0$, it follows that \mbox{${\bf h}_{\sf nuc}={\bf h}_{\sf int}$}, and the nuclear field is parallel to the $c$-axis. 
If the external field is then applied along the $c$-axis, the orientation of ${\bf h}_{\sf nuc}$ remains unchanged. If ${\bf h}_{\sf ext}$ is slowly turned on in the $ab$-plane then, as the magnitude is increased, ${\bf h}_{\sf nuc}$ will rotate by $\pi/2$ radians. Thus different strengths of the internal field will correspond to different angles $(\psi,\lambda)$ of ${\bf h}_{\sf nuc}$, even when the $(\theta,\phi)$ angles of the external field are kept constant.


For an external field applied in the $b$-direction the relaxation rate is given by,
\begin{align}
&\frac{1}{T_1(h_{\sf ext}^{b})} \approx  
(\mathcal{A}^{ca})^2  \ C_{st,1}\
 \Phi_{st,1}  \left( \frac{k_B T}{\Delta} \right)
 \frac{|{ \bf h}_{\sf ext}|^2}{|\langle { \bf h}_{\sf int} \rangle |^2+|{ \bf h}_{\sf ext}|^2} \nonumber \\
&\qquad + \frac{b_0^2(\mathcal{A}^{ba})^2}{v_b^2}   C_{st,k^2} \ \Phi_{st,k^2}  \left( \frac{k_B T}{\Delta} \right)
 \frac{|{ \bf h}_{\sf int}|^2}{|\langle { \bf h}_{\sf int} \rangle |^2+|{ \bf h}_{\sf ext}|^2} \nonumber \\
& \qquad + (\mathcal{A}^{aa})^2 C_{un,1} \ \Phi_{un,1} \left( \frac{k_B T}{\Delta}  \right) \nonumber \\
& \qquad + \frac{a_0^2(\mathcal{A}^{aa})^2}{v_a^2} C_{st,k^2} \ \Phi_{st,k^2}  \left( \frac{k_B T}{\Delta} \right),
\label{eq:Ba122hextvariationb}
\end{align}
where we emphasise the dependence on the magnitude of ${\bf h}_{\sf ext}$, and the constants 
$C_{st,1}$, $C_{un,1}$ and $C_{st,k^2}$ are defined in Eq.~(\ref{eq:Cst1}), Eq.~(\ref{eq:Cun1}) 
and Eq.~(\ref{eq:Cstksq}). 
If the external field is applied in the $a$-direction then,
\begin{align}
&\frac{1}{T_1(h_{\sf ext}^{a})} \approx  
(\mathcal{A}^{ca})^2  \ C_{st,1}\
 \Phi_{st,1}  \left( \frac{k_B T}{\Delta} \right)
 \frac{|{ \bf h}_{\sf ext}|^2}{|\langle { \bf h}_{\sf int} \rangle |^2+|{ \bf h}_{\sf ext}|^2} \nonumber \\
& \qquad + (\mathcal{A}^{aa})^2 C_{un,1} \ \Phi_{un,1} \left( \frac{k_B T}{\Delta}  \right)
 \frac{|{ \bf h}_{\sf int}|^2}{|\langle { \bf h}_{\sf int} \rangle |^2+|{ \bf h}_{\sf ext}|^2} \nonumber \\
& \qquad + \frac{a_0^2(\mathcal{A}^{aa})^2}{v_a^2} C_{st,k^2} \ \Phi_{st,k^2}  \left( \frac{k_B T}{\Delta} \right)
 \frac{|{ \bf h}_{\sf int}|^2}{|\langle { \bf h}_{\sf int} \rangle |^2+|{ \bf h}_{\sf ext}|^2} \nonumber \\
 &\qquad + \frac{b_0^2(\mathcal{A}^{ba})^2}{v_b^2}   C_{st,k^2} \ \Phi_{st,k^2}  \left( \frac{k_B T}{\Delta} \right).
\label{eq:Ba122hextvariationa}
\end{align}
In both cases the first term is the dominant contribution. 


An experimental measurement that either mapped out an octant of the ``doughnut'' shown 
in Fig.~(\ref{fig:Ba122doughnut}), or one that showed the dependence on external field magnitude 
illustrated in Fig.~(\ref{fig:Ba122hextmagnitude}), would provide strong support for the angle-resolved
theory developed in this paper.


\section{Quantitative determination of spin wave velocities from NMR}
\label{sec:characterisation}


Having established that it is possible to quantitatively match the theory of the NMR relaxation rate to 
experiment, we show how this can be exploited to make quantitative measurements of the dynamical 
properties of collinear anitferromagnets, with BaFe$_2$As$_2$ as an example.   
This technique could prove especially useful when crystal sizes are too small for inelastic neutron 
scattering measurements, as is often the case for newly synthesised materials.


NMR measurements of the internal field provide an excellent way of determining static properties 
of antiferromagnets, such as the ordering vector, the direction of the easy-axis and the energy scale 
of the resultant gap in the spin-wave spectrum. 
We have shown previously\cite{smerald10} how the gap, $\Delta$, can be measured in this way. 
Also, Knight shift measurements allow the components of the nuclear-electron coupling tensor 
to be determined\cite{kitagawa08}. 


We now argue that NMR relaxation rate measurements can be used to determine the 
hydrodynamic properties of collinear magnets, and in particular their low-energy 
spin wave velocities. 
We note that it has previously been suggested that it might be possible to place bounds on 
spin wave velocities in magnetic Fe pnictides from NMR experiments~\cite{ong09}.
However we believe that ours is the first theory sufficiently advanced to 
make a quantitative comparison with experiment.  

%
The first step is to determine the geometric combination $\bar{v}_s^3=v_av_bv_c$ by fitting data 
for the relaxation rate with external field in the  $ab$-plane using Eq.~(\ref{eq:T1hab}).


Next, we consider measuring the relaxation rate along the three principal crystal axes, and combining 
these in the linear combinations,
\begin{align} 
&\frac{1}{T_1^\pm} =
 \frac{1}{2} \left[  \frac{1}{T_1(h_{\sf ext}^c)} \pm\frac{1}{1-\sin^2\theta}\left( 
 \frac{1}{T_1(h_{\sf ext}^a)}  - \frac{1}{T_1(h_{\sf ext}^b) } \right) \right] . 
\end{align}  
The form factors associated with these combinations are constructed from Eq.~(\ref{eq:Ba122formfactor}), 
and given by,
\begin{align}
\mathcal{F}^+({\bf q})=
16(\mathcal{A}^{ba}s_as_b)^2,
\end{align}
in the case of $1/T_1^+$, and,
\begin{align}
\mathcal{F}^-({\bf q})=
16(\mathcal{A}^{aa}c_ac_b)^2,
\end{align}
for $1/T_1^-$, where $c_a$, $c_b$, $s_a$ and $s_b$ are given in Eq.~(\ref{eq:casa}).
It follows that,
\begin{align}
\frac{1}{T^+_1} \approx& 
\ C_{st,k^2}
 \left[  \frac{b_0\mathcal{A}^{ba}}{v_b}  \right]^2
\Phi_{st,k^2} \left(\frac{k_B T}{\Delta} \right),
\end{align}
and,
 \begin{align}
\frac{1}{T^-_1}& \approx 
(\mathcal{A}^{aa})^2 C_{un,1} \
\Phi_{un,1} \left( \frac{k_B T}{\Delta}  \right) \nonumber \\
&+C_{st,k^2}
 \left[ \frac{a_0\mathcal{A}^{aa}}{v_a}   \right]^2
\Phi_{st,k^2} \left(\frac{k_B T}{\Delta} \right),
\end{align}
where $|{\bf h}_{\sf ext}|\gg |\langle {\bf h}_{\sf int} \rangle|$ is assumed and  the 
constants $C_{st,1}$, $C_{un,1}$ and $C_{st,k^2}$ are defined in Eq.~(\ref{eq:Cst1}), 
Eq.~(\ref{eq:Cun1}) and Eq.~(\ref{eq:Cstksq}).
The only remaining unknowns are the spin wave velocities $v_a$ and $v_b$ and therefore these 
can be extracted from experimental data via,
\begin{align}
v_b=\sqrt{\frac{C_{st,k^2}\left[b_0 \mathcal{A}^{ba}\right]^2\Phi_{st,k^2}}{1/T^+_1}}.
\end{align}
and,
\begin{align}
v_a=\sqrt{\frac{C_{st,k^2}\left[a_0 \mathcal{A}^{aa}\right]^2
\Phi_{st,k^2}}{1/T^-_1-( \mathcal{A}^{aa})^2 C_{un,1}\Phi_{un,1} }},
\end{align}
The velocity $v_c$ then follows from the value of $\bar{v}_s$.


In order to perform a check on the values of the spin wave velocities one could also measure the 
dependence of $1/T_1(h^a_{\sf ext})$ and $1/T_1(h^b_{\sf ext})$ on $|{\bf h}_{\sf ext}|$ at 
constant temperature, as illustrated in Fig.~(\ref{fig:Ba122hextmagnitude}).  
Eq.~(\ref{eq:Ba122hextvariationb}) and Eq.~(\ref{eq:Ba122hextvariationa}) can be combined to give,
\begin{align}
\frac{1}{T_1(h_{\sf ext}^{a})} -&\frac{1}{T_1(h_{\sf ext}^{b})} \approx  
\left[  (\mathcal{A}^{aa})^2 C_{un,1} \ \Phi_{un,1} \left( \frac{k_B T}{\Delta} \right) \right. \nonumber \\
&  + \frac{a_0^2(\mathcal{A}^{aa})^2}{v_a^2} C_{st,k^2}  \Phi_{st,k^2}  \left( \frac{k_B T}{\Delta} \right) \nonumber \\
& \left. -  \frac{b_0^2(\mathcal{A}^{ba})^2}{v_b^2}   C_{st,k^2} \Phi_{st,k^2}  \left( \frac{k_B T}{\Delta} \right)
 \right]  \nonumber \\
&\times  \frac{|{ \bf h}_{\sf int}|^2}{|\langle { \bf h}_{\sf int} \rangle |^2+|{ \bf h}_{\sf ext}|^2},
\end{align}
where $v_a$ and $v_b$ are the only free parameters.


These techniques for measuring individual spin wave velocities involve combining relaxation rate 
measurements such that the dominant processes are canceled out, and the subleading terms 
are revealed. 
Thus a high degree of experimental accuracy is required. 
However, if a nuclear site can be found at which the internal field vanishes, then the fluctuations of the 
electron moments at the ordering vector are ``filtered out'' by the form factor for {\it all} orientations of 
internal field. 
In this case the above techniques are likely to become more powerful, since the cancellations required 
will be smaller.  
For example, this appears to be the case for the Y nucleus in YBa$_2$Cu$_3$O$_{6}$ 
[\onlinecite{unpublished}].


\section{Conclusion}
\label{sec:conclusions}


In this paper we presented a theory of NMR $1/T_1$ relaxation rates in a collinear 
antiferromagnet that provides quantitative fits to published data for $^{75}$As NMR
in BaFe$_2$As$_2$.  
All predictions are given in absolute units, and the spin fluctuations of electrons are
parameterized in terms of a small number of hydrodynamic parameters --- the ordered
moment $m_0$, transverse susceptibility $\chi_\perp$, anisotropy gap $\Delta$
and spin wave velocities $(v_a, v_b, v_c)$.  
The remaining parameters of the theory are the small number of matrix elements 
of the transferred hyperfine interaction between Fe electrons and the $^{75}$As nuclear spin.  
Since these can be determined from measurements of NMR spectra, the 
resulting theory has {\it no} adjustable parameters.  


A key feature of this theory --- and of $^{75}$As NMR on BaFe$_2$As$_2$ --- is a strong 
dependence of the $1/T_1$ relaxation rate on the orientation of the magnetic field.  
This angle-dependence can be traced back to the ``filtering'' of spin fluctuations by the 
form factor for transferred hyperfine interactions, which in turn depends on the orientation 
of the magnetic field.
Taking this into account, the theory correctly captures the qualitatively different temperature 
dependences of $1/T_1$ for $^{75}$As NMR in BaFe$_2$As$_2$ with field applied 
along the $[110]$ and $[001]$ directions~\cite{kitagawa08}.  


Moreover, since the theory is expressed only in terms of hydrodynamic parameters
of the magnetic electrons, this fitting procedure can be inverted, and angle-resolved $1/T_1$ 
measurements used to determine spin wave velocities directly from NMR experiments.
We have proposed a specific scheme for doing this from $^{75}$As NMR in BaFe$_2$As$_2$.  


While we have developed this theory with the particular goal of explaining $^{75}$As NMR 
experiments in BaFe$_2$As$_2$, the results have a much wider applicability.  
Firstly, a similar analysis can be applied to other collinear magnets, simply by modifying the form factor 
to take into account the symmetry environment of the nucleus in question.  
It appears that the ``doughnut'' shaped angle-dependence of $1/T_1$, shown in 
Fig.~(\ref{fig:Ba122doughnut}), remains valid for all nuclei that experience 
a non-zero internal magnetic field ${\bf h}_{\sf int}$.
The ``hole'' of the ``doughnut'' is aligned with ${\bf h}_{\sf int}$.  
For nuclei at high-symmetry sites where this internal field vanishes, the leading 
term in the relaxation rate is ``filtered out'' for {\it all} directions of external field, and 
the angular resolution acquires a more isotropic ``peanut''-like shape~\cite{unpublished}.


Although the theory developed in this paper is specific to an ordered antiferromagnet, 
the idea of angular resolution in $1/T_1$ measurements can easily be extended to the 
study of critical fluctuations.  
Relaxation rate data for BaFe$_2$As$_2$ with field applied in the $[110]$-direction 
show a significant upturn in $1/T_1$ as the (first-order) magnetic phase transition 
at $T_N= 135\ K$ is approached from the paramagnet [cf. Fig.~(\ref{fig:Ba122T1data})].  
This upturn occurs because, for field parallel to $[110]$, $1/T_1$ probes 
spin fluctuations near to the magnetic ordering vector ${\bf q}={\bf Q}$, and 
these are enhanced approaching the phase transition. 
In contrast, when the field is applied in the $[001]$ direction, there is no upturn in $1/T_1$.
For this field orientation, the form factor ``filters out'' critical fluctuations at ${\bf q}\approx{\bf Q}$, 
and $1/T_1$ is determined instead by spin fluctuations with ${\bf q}\approx0$.
Angle-resolved NMR experiments can therefore be used to isolate critical fluctuations
near to a phase transition in BaFe$_2$As$_2$ and other antiferromagnets\cite{unpublished}.

 
It can be seen in Fig.~(\ref{fig:Ba122T1data}) that the relaxation rate retains a significant angular dependence 
in the paramagnetic phase.
For an appreciable range of temperatures above $T_N$ the susceptibility will be dominated by fluctuations 
close to the ordering vector. 
Thus the idea of angle dependent ``filtering'' of spin fluctuations by the form factor will remain important.
This is a theme that we will develop in a separate publication\cite{unpublished}.
If the temperature is further increased such that $T\gg T_N, \Delta$ then the susceptibility will become isotropic 
in spin space and only weakly ${\bf q}$-dependent.
In this regime the angular resolution of the relaxation rate will be primarily due to the anisotropy 
of the nuclear-electron coupling tensor\cite{unpublished}.


This ability to tune between different spin fluctuations should also make angle-resolved 
NMR a powerful probe of unconventional magnetism, and in particular of exotic quantum 
phases in frustrated magnets.  
The absence of an ordered magnetic moment, lack of a large single crystal, or the requirement 
of large magnetic fields, often make these systems inaccessible to other probes, such 
as neutron scattering.  
One intriguing possibility is that angle-resolved $1/T_1$ measurements could provide a 
positive means of identifying the long-sought quantum spin-nematic state, a magnetic 
analogue of a liquid crystal, which does not break time reversal symmetry, and so does 
not give rise to magnetic Bragg peaks or static splitting in NMR 
spectra\cite{andreev84,chandra91,shannon06,sato09,svistov10,zhitomirsky10}.
This theme will be developed elsewhere\cite{inpreparation}.


{\it Acknowledgments.}   
We are particularly grateful to Kentaro Kitagawa and Masashi Takigawa for sharing their 
experimental data, and providing valuable advice at an early stage in this work.
We are also pleased to acknowledge useful discussions with Pietro Carretta and Frederic Mila.  
This work was supported under EPSRC grants EP/C539974/1 and EP/G031460/1


\appendix


\section{Calculation of the longitudinal suceptibility}
\label{App:susceptibility}


In this Appendix we calculate, at Gaussian order, the imaginary part of the dynamic susceptibility 
due to longitudinal fluctuations of the ordered moments. 
While the calculation is not original\cite{allen97}, we are not aware of any published derivation 
in absolute units, and therefore include it here for completeness.   
To make the ``filtering'' effect of the form factor more transparent, we choose to work in the 
Brillouin Zone associated with the orthorhombic lattice of magnetic sites in the paramagnet
(PMBZ), rather than the magnetic Brillouin Zone of the low temperature antiferromagnet.  


The dynamical susceptibility of the non-linear sigma model Eq.~(\ref{eq:nlsm}) can most
easily be calculated by considering the effect of an external field which varies in space and time,
\mbox{${\bf h}({\bf r}, \tau)$ [meV]}.   
The real space susceptibility [$\mu_B^2$] in the $(a,b,c)$ coordinate system of the crystal 
lattice is then given by,
\begin{align}
\left. \chi^{\alpha\beta}({\bf r},\tau) = -\left(g_l\mu_B\hbar V_{cell}\right)^2 \frac{\delta^2F}{\delta h^\alpha({\bf r},\tau) \delta h^\beta(0)} \right|_{{\bf h}=0}.
\label{eq:chifuncderiv}
\end{align} 
where $F=-\ln \mathcal{Z}$, and $\mathcal{Z}$ is the partition function, which, for the non-linear 
sigma model, is given by,
\begin{align}
\mathcal{Z} = \int \mathcal{D}{\bf n} \ \delta({\bf n}^2-1) \ e^{-\mathcal{S}[{\bf n},{\bf h}]}.
\end{align}


The longitudinal susceptibility can be accessed by considering both staggered and uniform fields applied parallel 
to the ordering axis.
We first consider the effect of a staggered field, ${\bf h}_{st}$.  
This couples directly to the antiferromagnetic order parameter,~${\bf n}$, according to   
\begin{align}
\mathcal{S}[{\bf n},{\bf h}_{st}]  = \mathcal{S}[{\bf n}] - \frac{S}{\hbar V_{cell} }\int d^3r d\tau \ 
{\bf n}.{\bf h}_{st},
\label{eq:Shs}
\end{align}
where $S$ is the total spin per site. 
The order parameter field can be parametrised as 
\mbox{${\bf n}=\left(\sqrt{1-\phi_1^2-\phi_2^2}, \phi_1,\phi_2\right)$}, and, for temperatures at 
which the fluctuations around the ordered state are small, $\phi_1,\phi_2 \ll 1$.   
To Gaussian order in $\phi_1,\phi_2$, Eq.~(\ref{eq:nlsm}) becomes,
 \begin{align}
\mathcal{S}[\boldsymbol{\phi}] = &\frac{1}{2\hbar V_{cell}}\int d^3r d\tau 
\left[  
\hbar^2\chi_\perp (\partial_\tau \boldsymbol{\phi})^2 + \sum_\alpha \rho_\alpha (\partial_\alpha \boldsymbol{\phi})^2  \right. \nonumber \\
& \left. 
\qquad \qquad \qquad \qquad  \qquad +\chi_\perp \Delta^2 \boldsymbol{\phi}^2  
 \right],
 \end{align}
where $\boldsymbol{\phi}=(\phi_1,\phi_2)$.   
Likewise Eq.~(\ref{eq:Shs}) for the staggered field gives,
\begin{align}
\mathcal{S}[\boldsymbol{\phi}, h_{st}]  = \mathcal{S}[\boldsymbol{\phi}] -\frac{Sh_{st}}{\hbar V_{cell}}\int d^3r d\tau&  
\left( 1-\frac{\boldsymbol{\phi}^2}{2}  \right).
\end{align}
Starting from the partition function,
\begin{align}
\mathcal{Z}_{st} = \int \mathcal{D}\boldsymbol{\phi}  \ e^{-\mathcal{S}[\boldsymbol{\phi}, h_{st}]},
\end{align}
and calculating the longitudinal, staggered susceptibility from Eq.~(\ref{eq:chifuncderiv}) gives,
\begin{align}
 \chi_{\parallel,{st}}({\bf r},\tau) = \eta({\bf r})   \left( \frac{g_l\mu_B S}{2} \right)^2  &
  \left( \left\langle \boldsymbol{\phi}^2({\bf r},\tau) \boldsymbol{\phi}^2(0) \right\rangle  \right. \nonumber \\
&\left.   -  \left\langle \boldsymbol{\phi}^2({\bf r},\tau) \right\rangle \left\langle \boldsymbol{\phi}^2(0) \right\rangle  \right),
  \label{eq:suceprealspace}
   \end{align}
where $\eta({\bf r})$ accounts for the staggered nature of the field and, 
  \begin{align}
 \left\langle \mathcal{O} \right\rangle =   \frac{1}{\mathcal{Z}}\int\mathcal{D}\boldsymbol{\phi} \ \mathcal{O} \ e^{-\mathcal{S}[\boldsymbol{\phi}]}.  
 \end{align}


The susceptibility in Eq.~(\ref{eq:T1suscep}) is the Fourier transform of that entering Eq.~(\ref{eq:suceprealspace}). We choose to work in the full PMBZ in reciprocal space, rather than the reduced magnetic Brillouin Zone (MBZ), since this makes the physical picture of the interaction with the form factor clearer. For two sublattice antiferromagnetic order there are two identical cones of spin wave excitations in the PMBZ, one at ${\bf q}=0$ and the other at the ordering vector ${\bf q}={\bf Q}$. After Fourier transform, the fields $\phi_1$ and $\phi_2$ describe two independent cones of bosonic excitation. The field Fourier transforms are defined by,
\begin{align}
&\phi_1({\bf r},\tau)=\frac{1}{\beta \hbar}\sum_{i\omega_n} \frac{V_{cell}}{(2\pi)^3} \int d^3q \ e^{i({\bf q}.{\bf r}+\omega_n \tau)}\phi_1({\bf q},i\omega_n) \nonumber \\
&\phi_2({\bf r},\tau)=\frac{1}{\beta \hbar}\sum_{i\omega_n}  \frac{V_{cell}}{(2\pi)^3} \int d^3q \ e^{i(({\bf q}-{\bf Q}).{\bf r}+\omega_n \tau)}\phi_2({\bf q},i\omega_n),
\label{eq:FT}
 \end{align}
where $\phi_1({\bf q},i\omega_n)$  describes the excitation cone at ${\bf q}=0$ and  $\phi_2({\bf q},i\omega_n)$ describes the excitation cone at ${\bf q}={\bf Q}$.
It follows from substituting the field Fourier transforms into the action that the two field averages are,
\begin{align}
&\langle  \phi_1({\bf q},i\omega_n)\phi_1({\bf q}^\prime,i\omega_n^\prime)  \rangle = \nonumber \\
& \qquad \qquad \delta({\bf q}+{\bf q}^\prime)\delta(i\omega_n+i\omega_n^\prime)G_0^{\phi_1}({\bf q},i\omega_n) \nonumber \\
&\langle  \phi_2({\bf q},i\omega_n)\phi_2({\bf q}^\prime,i\omega_n^\prime) \rangle =  \nonumber \\
& \qquad \qquad \delta({\bf q}+{\bf q}^\prime-2{\bf Q})\delta(i\omega_n+i\omega_n^\prime)G_0^{\phi_2}({\bf q},i\omega_n) \nonumber \\
&\langle  \phi_1({\bf q},i\omega_n)\phi_2({\bf q}^\prime,i\omega_n^\prime)  \rangle = 0,
\label{eq:Gfunc}
 \end{align}
 with,
\begin{align}
G_0^{\phi_1}({\bf q},i\omega_n) &= \frac{1}{\hbar\chi_\perp} \frac{1}{\omega_{1,{\bf q}}^2  -  (i\omega_n)^2} \nonumber \\
G_0^{\phi_2}({\bf q},i\omega_n) &= \frac{1}{\hbar\chi_\perp} \frac{1}{\omega_{2,{\bf q}}^2  -  (i\omega_n)^2}.
\label{eq:Gfunc2}
 \end{align}
The energies,
\begin{align}
&\hbar \omega_{1,{\bf q}} = \sqrt{\Delta^2+\sum_\alpha v_\alpha^2 q_\alpha^2} \nonumber \\
&\hbar \omega_{2,{\bf q}} = \sqrt{\Delta^2+\sum_\alpha v_\alpha^2 (q_\alpha-Q_\alpha)^2}, 
\label{eqapp:dispersion}
 \end{align}
describe the dispersion of the two spin wave cones and,
 \begin{align}
v_\alpha=\sqrt{ \frac{\rho_\alpha}{\chi_\perp} },
\label{eq:velocities}
 \end{align}
gives the spin wave velocities [meV$\AA$], where \mbox{$\alpha=\{ a,b,c \}$}. 


Expanding four field averages using Wick's theorem, using Eq.~(\ref{eq:Gfunc}) to substitute for the two field averages, Fourier transforming the fields using Eq.~(\ref{eq:FT}) and rewriting the staggering parameter as $\eta({\bf r})=e^{\pm i{\bf Q}. {\bf r}}$, gives,
\begin{align}
&    \chi_{\parallel,{st}}({\bf q},i\omega_n)  \approx
\left( \frac{g_l \mu_B S}{2} \right)^2 V_{cell}\int \frac{d^3k}{(2\pi)^3} \frac{1}{\beta\hbar} \sum_{i\nu_n} 
 \nonumber \\
&   \frac{2}{(\chi_\perp\hbar)^2} \left( \frac{1}{\omega^2_{1,{\bf k}}-(i\nu_n)^2} \ \frac{1}{\omega^2_{1,{\bf k}+{\bf q}-{\bf Q}}-(i\nu_n+i\omega_n)^2} \right. \nonumber \\
& \left.+ \frac{1}{\omega^2_{2,{\bf k}}-(i\nu_n)^2} \ \frac{1}{\omega^2_{2,{\bf k}+{\bf q}+{\bf Q}}-(i\nu_n+i\omega_n)^2} \right), \nonumber \\
\label{eq:suscmatsubarasums}
\end{align}
where $g_l$ is the Land\'{e} g-factor. 
Performing the  Matsubara sums over $i\nu_n$ and analytically continuing to real frequencies we find
\begin{align}
&\Im m \left\{\chi_{\parallel,st}({\bf q},\omega_0) \right\} \approx 
 \left(\frac{g_l\mu_BS}{2} \right)^2 \frac{\pi}{2} \hbar V_{cell}  \frac{1}{\chi_\perp^2} 
 \nonumber \\
&  \left\lgroup \int_{{\bf k}\approx 0} \frac{d^3k}{(2\pi)^3}  
 \frac{n_B(\omega_{1,{\bf k}})-n_B(\omega_{2,{\bf k}+{\bf q}})}{\hbar\omega_{1,{\bf k}} \ \hbar\omega_{2,{\bf k}+{\bf q}}} \right. \nonumber \\
&\times \left[ \delta(\hbar\omega_{1,{\bf k}}-\hbar\omega_{2,{\bf k}+{\bf q}}+\omega_0) 
-\delta(\hbar\omega_{2,{\bf k}+{\bf q}}-\hbar\omega_{1,{\bf k}}+\omega_0) \right] \nonumber \\
&    + \int_{{\bf k}\approx {\bf Q}} \frac{d^3k}{(2\pi)^3}
\frac{n_B(\omega_{2,{\bf k}})-n_B(\omega_{1,{\bf k}+{\bf q}})}{\hbar\omega_{2,{\bf k}} \ \hbar\omega_{1,{\bf k}+{\bf q}}}  \nonumber \\
&\left. \times \left[ \delta(\hbar\omega_{2,{\bf k}}-\hbar\omega_{1,{\bf k}+{\bf q}}+\omega_0) 
-\delta(\hbar\omega_{1,{\bf k}+{\bf q}}-\hbar\omega_{2,{\bf k}}+\omega_0) \right]  \right\rgroup, 
\label{eqapp:stagsuscep}
\end{align}
where $\omega_{1,{\bf q}}= \omega_{2,{\bf q}+{\bf Q}}$ has been used and,
\begin{align}
n_B(\omega_{\bf q}) = \frac{1}{e^{\hbar\omega_{\bf q}/k_BT}-1},
\end{align}
is the standard Bose factor.
The staggered susceptibility describes scattering of spin waves between two cones of excitations 
separated by wavevector ${\bf q}= {\bf Q}$.  
It is peaked at ${\bf q}= {\bf Q}$, and the sharpness of the peak increases with decreasing temperature.  
This is because, at low temperatures, less of the excitation cone is accessible to the spin-wave fluctuations. 


There is also a contribution to the susceptibility associated with the application of a uniform field, ${\bf h}_{un}$. 
The coupling between the order parameter field and ${\bf h}_{un}$ can be included in the action 
as\cite{allen97,benfatto06},
\begin{align}&
\mathcal{S}[{\bf n},{\bf h}_{un}]  =\mathcal{S}[{\bf n}]- \frac{1}{V_{cell}}\int d^3r d\tau
\left[  
i\chi_\perp {\bf h}_{un}.({\bf n} \times \partial_\tau {\bf n}) \right. \nonumber \\
& \qquad \left.+ \frac{\chi_\perp}{2\hbar}\left( {\bf h}_{un}^2- ({\bf n}.{\bf h}_{un})^2 \right)
-\frac{S}{4\hbar}\sum_{\alpha} \alpha \partial_\alpha ({\bf n}.{\bf h}_{un})
 \right].
\end{align}
Performing an expansion in the fields $\boldsymbol{\phi}$, as before, leads to
  \begin{align}
&\mathcal{S}[\boldsymbol{\phi}, h_{un}]  =\mathcal{S}[\boldsymbol{\phi}] - \frac{1}{V_{cell}}  \int  d^3r d\tau \nonumber \\
&\qquad \left[  
i\chi_\perp h_{un}  \left( \phi_1\partial_\tau \phi_2 -\phi_2\partial_\tau \phi_1 \right)  -\frac{\chi_\perp h_{un}^2}{2\hbar}(\phi_1^2+\phi_2^2)
 \right].
\end{align} 
After repeating a similar set of manipulations to above, the contribution to the imaginary 
part of the susceptibility is found to be,
\begin{align}
& \Im m \left\{ \chi_{\parallel,un}({\bf q},\omega_0) \right\}  \approx
(g_l\mu_B)^2  \frac{\pi}{2} \hbar V_{cell}  
 \nonumber \\
&  \left\lgroup \int_{{\bf k}\approx 0} \frac{d^3k}{(2\pi)^3}  
\left( n_B(\omega_{1,{\bf k}})-n_B(\omega_{1,{\bf k}+{\bf q}}) \right) \right. \nonumber \\
&\times \left[ \delta(\hbar\omega_{1,{\bf k}}-\hbar\omega_{1,{\bf k}+{\bf q}}+\omega_0) 
-\delta(\hbar\omega_{1,{\bf k}+{\bf q}}-\hbar\omega_{1,{\bf k}}+\omega_0) \right] \nonumber \\
&    + \int_{{\bf k}\approx {\bf Q}} \frac{d^3k}{(2\pi)^3}
\left( n_B(\omega_{2,{\bf k}})-n_B(\omega_{2,{\bf k}+{\bf q}}) \right)  \nonumber \\
&\left. \times \left[ \delta(\hbar\omega_{2,{\bf k}}-\hbar\omega_{2,{\bf k}+{\bf q}}+\omega_0) 
-\delta(\hbar\omega_{2,{\bf k}+{\bf q}}-\hbar\omega_{2,{\bf k}}+\omega_0) \right]  \right\rgroup
\label{eqapp:unsuscep}
  \end{align}
This describes scattering of spin waves within an excitation cone, and is peaked at scattering 
vector ${\bf q}=0$.  
As with the staggered susceptibility, the peak becomes sharper at lower temperatures, where 
less of the spin wave cone can be accessed. 


The imaginary part of the susceptibility is given by the sum of the two terms, 
\begin{align}
\Im m \left\{ \chi_{\parallel}({\bf q},\omega_0) \right\} = &
\Im m \left\{ \chi_{\parallel,st}({\bf q}\approx {\bf Q},\omega_0) \right\} \nonumber \\
&+\Im m \left\{ \chi_{\parallel,un}({\bf q}\approx 0,\omega_0) \right\}.
  \end{align}
These results are used in Section~\ref{subsec:suscep} of the paper.


\subsection{Spin wave theory from Heisenberg model}


Since experiments on magnetic Fe pnictides are often discussed in terms of a Heisenberg 
model\cite{yildrim08,si08}, we sketch below an equivalent calculation of the longitudinal susceptibility 
within conventional spin wave theory.
We stress that the non-linear sigma model correctly reproduces the low energy behaviour 
of the Heisenberg antiferromagnet, as it does for any microscopic model with the correct symmetries. 
We provide in Table~\ref{tab:heisenbergdictionary} a dictionary to translate between these two 
models at the Gaussian level of approximation.


The Heisenberg Hamiltonian with the correct symmetries for the magnetically ordered phase of 
BaFe$_2$As$_2$ is\cite{ewings08},
\begin{align}
&\mathcal{H}=J_{1a}\sum_{\langle ij \rangle_{1a}}\textsf{S}_i.\textsf{S}_j
+J_{1b}\sum_{\langle ij \rangle_{1b}}\textsf{S}_i.\textsf{S}_j
 +J_{1c}\sum_{\langle ij \rangle_{1c}}\textsf{S}_i.\textsf{S}_j
\nonumber \\
&+J_2\sum_{\langle ij \rangle_{2}} \textsf{S}_i.\textsf{S}_j 
-K_{ab} \sum_i  \left((\textsf{S}_i^a)^2-(\textsf{S}_i^b)^2\right) + K_c \sum_i (\textsf{S}_i^c)^2,
\label{eq:Ham}
\end{align}
where $\langle ij \rangle_{1\alpha}$ counts first-neighbour bonds in the $\alpha$-direction, 
$\langle ij \rangle_2$ second-neighbour bonds in the $a$-$b$ plane, and $K_{ab}$ and 
$K_{c}$ are single-ion anisotropies.  


We consider the case $K_{ab}=K_{c}=K$, and use linear spin wave theory to calculate 
the spin wave dispersion and longitudinal susceptibility.  
Due to the large ($\Delta \approx 100\ K$) gap in the spin-wave spectrum, $1/S$ corrections 
are small and can be safely neglected.
By rewriting the spin degrees of freedom in terms of Holstein-Primakoff bosons\cite{fazekas} 
one can transform the Hamiltonian to,
\begin{align}
H_{lsw}=\frac{1}{2} \sum_{{\bf k}\in PMBZ} 
\left(  a^\dagger_{\bf k},a^{\phantom\dagger}_{-{\bf k}} \right)
 \left(
 \begin{array}{cc}
A_{\bf k} & B_{\bf k} \\
B_{\bf k}  & A_{\bf k}
\end{array}
 \right)
  \left(
 \begin{array}{c}
a^{\phantom\dagger}_{\bf k} \\
a^\dagger_{-{\bf k}}
\end{array}
 \right)
 -A_{\bf k},
\end{align}
where,
\begin{align}
A_{\bf k}&=2S(2J_2+J_{1a}-J_{1b}(1-\cos k_b)+J_{1c})+4SK \nonumber \\
B_{\bf k}&=2S(2J_2\cos k_a \cos k_b+J_{1a}\cos k_c+J_{1c} \cos k_c).
\end{align}
Performing a Bogoliubov transformation with the coherence factors,
\begin{align}
&u_{\bf k}= \frac{1}{\sqrt{2}} \sqrt{ \frac{A_{\bf k}}{\omega_{\bf k}}+1 }, \qquad
v_{\bf k}= \frac{1}{\sqrt{2}} \sqrt{ \frac{A_{\bf k}}{\omega_{\bf k}}-1 }
\end{align}
results in,
\begin{align}
\omega_{\bf k}\approx\sqrt{A_{\bf k}^2-B_{\bf k}^2}.
\end{align}
It follows that the imaginary part of the longitudinal susceptibility is,
\begin{align}
&\Im m  \left\{ \chi_\parallel({\bf q},\omega_0) \right\}
\approx \frac{\pi}{2}\frac{(g\mu_B)^2   \hbar}{N}  \nonumber \\
& \sum_{{\bf k}\in PMBZ} \hspace{-5mm} \left( u_{\bf k} u_{{\bf k}+{\bf q}-{\bf Q}} + v_{\bf k} v_{{\bf k}+{\bf q}-{\bf Q}}  \right)^2
(n_B( \omega_{\bf k})-n_B( \omega_{{\bf k}+{\bf q}-{\bf Q}})) \nonumber \\
& \left[ \delta(  \hbar \omega_{\bf k}- \hbar \omega_{{\bf k}+{\bf q}-{\bf Q}} +\omega_0) -  \delta(  \hbar \omega_{{\bf k}+{\bf q}-{\bf Q}}- \hbar \omega_{\bf k} +\omega_0)  \right].
\end{align} 


The combination of coherence factors appearing in the expression for the susceptibility can be 
rewritten in terms of the parameters of the non-linear sigma model according to,
\begin{align}
& \left( u_{\bf k} u_{{\bf k}+{\bf q}-{\bf Q}} + v_{\bf k} v_{{\bf k}+{\bf q}-{\bf Q}}  \right)^2  \approx  \nonumber \\
&\qquad \qquad \qquad \left\{
 \begin{array}{cc}
\frac{1}{\omega_{\bf k} \omega_{ {\bf k}+{\bf q}-{\bf Q} }  } \left( \frac{S}{2\chi_\perp} \right)^2, & {\bf q}\approx {\bf Q} \\
1, & {\bf q}\approx 0
\end{array}
\right.
\end{align}
with,
\begin{align}
\omega_{\bf k}\approx \sqrt{\Delta^2+\sum_\alpha v_\alpha^2k_\alpha^2}.
\end{align}
The relationships between these hydrodynamic parameters and the exchange integrals 
of the Heisenberg Hamiltonian are shown in Table~\ref{tab:heisenbergdictionary}.
\begin{table}
\begin{center}
\footnotesize
  \begin{tabular}{| c | c |}
    \hline
non-linear  & Heisenberg   \\ 
sigma model  & model  \\ \hline 
\multirow{2}{*}{$\Delta$} & \multirow{2}{*}{$4S[K(J_{1a}+2J_2+J_{1c})+K^2]^{\frac{1}{2}}$} \\
&\\ \hline
\multirow{2}{*} {$v_a$} & \multirow{2}{*}{$2S[4J_2^2-4J_{1a}J_2+J_{1a}^2+2J_2J_{1c}-J_{1a}J_{1c}]^{\frac{1}{2}}$} \\
&  \\ \hline
\multirow{2}{*}{$v_b$} & $2S [4J_2^2-2J_{1a}J_2+2J_{1b}J_2+2J_2J_{1c}$  \\
& $\qquad \quad+J_{1a}J_{1b}+J_{1b}J_{1c}-2J_{1b}^2+2KJ_{1b}]^{\frac{1}{2}}$  \\ \hline
\multirow{2}{*}{$v_c$} & \multirow{2}{*}{$2S[2J_2J_{1c}-J_1J_{1c}+J_{1c}^2]^{\frac{1}{2}} $} \\
&  \\ \hline
\multirow{2}{*}{$\chi_\perp$} & \multirow{2}{*}{$1/(4J_{1a}+8J_2+4J_\perp+4K)$} \\
& \\ \hline
    \end{tabular}
\end{center} 
\caption{\footnotesize{
Relationship between the hydrodynamic parameters which characterize the non-linear 
sigma model and the exchange integrals which enter the Heisenberg Hamiltonian.
Both models are treated at the same level of approximation : 
linear spin wave theory for the Heisenberg model and a 
Gaussian mean-field approximation for the non-linear sigma model.
}}
\label{tab:heisenbergdictionary}
\end{table}
%


\section{Spectral representations of form factors}
\label{App:dos}


The simplest way to perform the momentum integrals that occur in Section~\ref{subsec:abfield} 
and Section~\ref{subsec:cfield} is to make a transformation from momentum to energy space. 
This is done via the density of states,
\begin{align}
g(\epsilon)= \int_{cone} \frac{d^3k}{(2\pi)^3} \delta(\epsilon-\epsilon_{\bf k}),
\end{align}
with,
\begin{align}
\epsilon_{\bf k}=\sqrt{\Delta^2+k^2}.
\end{align}
The integration region is spherically symmetric and so it is natural to use polar coordinates.
Using,
\begin{align}
 \delta(\epsilon-\sqrt{\Delta^2+ k^2}) = 
\frac{\epsilon \ \delta(k-\sqrt{\epsilon^2-\Delta^2})}{\sqrt{\epsilon^2-\Delta^2}}.
\end{align}
 leads to a density of states, 
\begin{align}
g(\epsilon)&=4\pi \int_{cone} \frac{dk}{(2\pi)^3} k^2 \frac{\epsilon \ \delta(k-\sqrt{\epsilon^2-\Delta^2})}{\sqrt{\epsilon^2-\Delta^2}}  \nonumber \\
&=\frac{1}{2\pi^2} \epsilon\sqrt{\epsilon^2-\Delta^2}.
\end{align}


The expression for the relaxation rate with field parallel to $[001]$, Eq.~(\ref{eq:T1ksq}), 
contains additional factors of $k_a^2$ and $k_b^2$ in the integrand. 
These are most easily handled using the spectral representation,
\begin{align}
A_{k_a}(\epsilon)=\int_{cone} \frac{d^3k}{(2\pi)^3} k_a^2 \delta(\epsilon-\epsilon_{\bf k}) .
\end{align}
This is calculated as,
\begin{align}
A_{k_a}(\epsilon)&= \int \frac{d\Omega}{(2\pi)^3} \sin^2\theta \cos^2\phi \int dk k^4 \frac{\epsilon \ \delta(k-\sqrt{\epsilon^2-\Delta^2})}{\sqrt{\epsilon^2-\Delta^2}} \nonumber \\
&= \frac{1}{6\pi^2} \epsilon(\epsilon^2-\Delta^2)^{\frac{3}{2}}.
\end{align}


By a completely analogous method,
\begin{align}
A_{k_b}(\epsilon)&=\int_{cone} \frac{d^3k}{(2\pi)^3} k_b^2 \delta(\epsilon-\epsilon_{\bf k}) \nonumber \\
&=  \frac{1}{6\pi^2} \epsilon(\epsilon^2-\Delta^2)^{\frac{3}{2}}.
\end{align}


\section{Integrating products of Bose functions}
\label{App:boseintegrals}


The calculation of the relaxation rate requires integration of products of polynomials and Bose functions. We show here the results that we make use of in Section~\ref{subsec:abfield} and Section~\ref{subsec:cfield}. 
For integer n,
\begin{align}
\int\mathrm{d}x \ \frac{x^n\mathrm{e}^x}{(\mathrm{e}^x-1)^2}=-\sum_{m=0}^n  \frac{n!}{(n-m)!} x^{n-m} \mathrm{Li}_m\left[\mathrm{e}^{-x}\right],
\end{align}
where,
\begin{align}
\mathrm{Li}_m(z) = \sum_{l=0}^\infty z^l/l^m.
\end{align}
The integrals required are,
\begin{align}
&\int_{\frac{\Delta}{T}}^\infty \mathrm{d}x \ \frac{\mathrm{e}^x}{(\mathrm{e}^x-1)^2}=\mathrm{Li}_0 \left[\mathrm{e}^{-\frac{\Delta}{T}}\right]=\frac{1}{\mathrm{e}^{\frac{\Delta}{T}}-1}  \\
&\int_{\frac{\Delta}{T}}^\infty \mathrm{d}x \ \frac{x^2\mathrm{e}^x}{(\mathrm{e}^x-1)^2}=\frac{\Delta^2}{T^2}\mathrm{Li}_0 \left[\mathrm{e}^{-\frac{\Delta}{T}}\right]\nonumber \\
& \qquad \qquad +2\frac{\Delta}{T}\mathrm{Li}_1 \left[\mathrm{e}^{-\frac{\Delta}{T}}\right]+2\mathrm{Li}_2 \left[\mathrm{e}^{-\frac{\Delta}{T}}\right]  \\
&\int_{\frac{\Delta}{T}}^\infty \mathrm{d}x \ \frac{x^4\mathrm{e}^x}{(\mathrm{e}^x-1)^2}=\frac{\Delta^4}{T^4}\mathrm{Li}_0 \left[\mathrm{e}^{-\frac{\Delta}{T}}\right]+4\frac{\Delta^3}{T^3}\mathrm{Li}_1 \left[\mathrm{e}^{-\frac{\Delta}{T}}\right] \nonumber \\
&+12\frac{\Delta^2}{T^2}\mathrm{Li}_2 \left[\mathrm{e}^{-\frac{\Delta}{T}}\right] 
+24\frac{\Delta}{T}\mathrm{Li}_3 \left[\mathrm{e}^{-\frac{\Delta}{T}}\right] +24\mathrm{Li}_4 \left[\mathrm{e}^{-\frac{\Delta}{T}}\right]. 
\end{align}



\end{document}